\documentclass[aps,prd,twocolumn,floatfix]{revtex4-2}
\usepackage{graphicx}
\usepackage{amstext}
\usepackage{array}  
\newcolumntype{L}{>{$}l<{$}}
\usepackage{pstricks}
\usepackage{color}
\usepackage{placeins}

\usepackage{booktabs, tabularx, array}
\usepackage{soul}
\usepackage[colorinlistoftodos, textsize=small]{todonotes}
\usepackage{pdfcomment}

\usepackage[binary-units=true,separate-uncertainty=true]{siunitx}
\usepackage{lineno,hyperref}
\usepackage{tikz}
\usepackage{amssymb,amsmath,mathtools}
\usepackage{caption}
\usepackage{xspace}
\usepackage{booktabs}
\usepackage{textcomp}
\usepackage[frozencache]{minted}
\usepackage[percent]{overpic}
\usepackage[list=true,listformat=simple,margin=10pt]{subcaption}
\usepackage{tikz-feynman}
\usepackage{multirow}
\usepackage{tocloft}
\usepackage{paracol}

\usepackage{array}
\newcommand{\ttHa}{$t\bar{t}H\gamma$}
\newcommand{\tHa}{$tH\gamma$}
\newcommand{\MET}{$\vec{\mspace{2mu}\mathrm{p}}_{\mathrm{T}}^{\mathrm{miss}}$}
\begin{document}

\title{Feasibility study of single top-quark and top-quark pair production in association with a Higgs Boson and a Photon at the LHC}

\author{Ashfaq~Ahmad}
\thanks{ashfaq.ahmad@cern.ch}
\author{Kamran~Ahmad}
\thanks{kamran.ahmad@cern.ch}
\author{Shoaib~Ahmad~Khan}
\affiliation{National Centre for Physics, Shahdra Valley Road, Islamabad, Pakistan}

\begin{abstract}
A feasibility study for the Standard Model Higgs boson produced in association with a single top-quark or a top-quark pair and a photon (\tHa\, and \ttHa) is presented, using simulated pp collision data corresponding to an integrated luminosity of 350\,fb$^{-1}$ at $\sqrt{s}=13.6$\,TeV. This study was conducted using simulated data generated with MadGraph. The study was performed in the context of the CMS experiment, where detector effects were incorporated using Delphes. Final states are selected through the leptonic decay of W boson, and the Higgs boson decays to two b-quarks. Signal events are separated from background events using multivariate techniques such as Boosted Decision Trees\,(BDT). The expected cross section for single top-quark production in association with a Higgs boson and a photon, $\sigma$(\tHa), was measured to be 1.31 fb at 13.6\,TeV, with a significance of 7.5 standard deviations from the background-only hypothesis for a luminosity of 350\,fb$^{-1}$ at 13.6\,TeV. The expected cross section for $\sigma$(\ttHa) was measured to be $2.94^{+0.196}_{-0.276}$ fb at 13.6\,TeV, with a significance of 6.6 standard deviations from the background-only hypothesis for a luminosity of 350\,fb$^{-1}$ at 13.6\,TeV. Importantly, both the \tHa\, and \ttHa\, processes are already feasible with the currently accumulated Run 3 dataset, demonstrating strong potential for early measurements.
\end{abstract}


\maketitle

\section{Introduction}
\label{sec:intro}
Since the CMS and ATLAS Collaborations announced the discovery of a new boson in 2012\,\cite{Higgs_CMS,Higgs_ATLAS}, the aim of the experimental studies is to assess how well its properties match those of the Standard Model\,(SM) Higgs boson\,\cite{Gauge_Vector_Mesons,Gauge_Fields,Gauge_Bosons,Massless_Particles,SSB,SB_non_abelian}. Thus far, the observations align with SM predictions within experimental uncertainties\,\cite{Higgs_uncer_1,Higgs_uncer_2,Higgs_uncer_3,Higgs_uncer_4}. The presence of experimental uncertainties mean that there sill remains a plenty of room for new physics in Higgs sector\,\cite{Higgs,Higgs_NP}.

\begin{figure*}[tbp] 
	\centering 
	\begin{minipage}{0.335\textwidth}
		\centering
		\resizebox{1\textwidth}{!}{ 
			\begin{tikzpicture}
				\begin{feynman}
					\vertex (a);
					\vertex [right=of a, xshift=-0.9cm] (b);
					\vertex [above right=of b, xshift=-0.7cm, yshift=-0.5cm] (d1) {\scriptsize $H$};
					\vertex [below right=of b, xshift=-0.65cm, yshift=0.8cm] (d2);
					
					\vertex [above right=of d2, xshift=-0.65cm, yshift=-0.49cm] (d21) {\scriptsize $\gamma$};
					\vertex [below right=of d2, xshift=-0.5cm, yshift=0.65cm] (d22);
					
					\vertex [above right=of d22, xshift=-0.5cm, yshift=-0.8cm] (d221);
					\vertex [below right=of d22, xshift=-0.9cm, yshift=0.69cm] (d222) {\scriptsize $b$};
					
					\vertex [above right=of d221, xshift=-0.8cm, yshift=-0.7cm] (d2211) {\scriptsize $\nu_l$};
					\vertex [below right=of d221, xshift=-0.65cm, yshift=1.1cm] (d2212) {\scriptsize $l^+$};
					
					\vertex [above left=of a, xshift=0.55cm, yshift=-0.1cm] (a1);
					\vertex [below left=of a, xshift=0.55cm, yshift=0.1cm] (a2);
					\vertex [above left=of a1, xshift=-0.1cm,yshift=-0.6cm] (a11) {\scriptsize $u$};
					\vertex [above right=of a1, xshift=0.15cm, yshift=-0.6cm] (b1);

					\vertex [below left=of a2, xshift=-0.1cm, yshift=0.6cm] (a21) {\scriptsize $g$};
					\vertex [below right=of a2, xshift=0.15cm, yshift=0.6cm] (b2);

					\diagram {
						(a) -- [fermion, line width = 0.02cm, arrow size = 0.03cm, edge label = \scriptsize $t$] (b);
						(d1) -- [scalar, line width = 0.02cm, arrow size = 0.03cm] (b) -- [fermion, line width = 0.02cm, arrow size = 0.03cm , edge label = \scriptsize $t$] (d2);
						(a2) -- [fermion, line width = 0.02cm, arrow size = 0.03cm, edge label = \scriptsize $b$] (a) -- [boson, line width = 0.02cm, arrow size = 0.03cm, edge label = \scriptsize $W^{-}$] (a1);
						(a11) -- [fermion, line width = 0.02cm, arrow size = 0.03cm] (a1);
						(a2) -- [gluon, line width = 0.02cm] (a21);
						(a1) -- [fermion, line width = 0.02cm, arrow size = 0.03cm, edge label = \scriptsize $d$] (b1);
						(b2) -- [fermion, line width = 0.02cm, arrow size = 0.03cm, edge label = \scriptsize $\bar{b}$] (a2);

						(d2) -- [boson, line width = 0.02cm] (d21);
						(d2) -- [fermion, line width = 0.02cm, arrow size = 0.03cm, edge label = \scriptsize $t$] (d22);
						
						(d22) -- [boson, line width = 0.02cm, edge label' = \scriptsize $W^{+}$] (d221);
						(d22) -- [fermion, line width = 0.02cm, arrow size = 0.03cm] (d222);

						(d2212) -- [fermion, line width = 0.02cm, arrow size = 0.03cm] (d221) -- [fermion, line width = 0.02cm, arrow size = 0.03cm] (d2211)
					};

				\end{feynman}
			\end{tikzpicture}
		}
	\end{minipage}
	\hspace{0.3cm}
	\begin{minipage}{0.335\textwidth} 
		\centering
		\resizebox{0.82\textwidth}{!}{
			\begin{tikzpicture}
				\begin{feynman}
					\vertex (a);
					\vertex [right=of a] (b);
					\vertex [above right=of b, xshift=-0.6cm, yshift=-0.99cm] (d1) {\scriptsize$b$};
					\vertex [below right=of b, xshift=-0.6cm, yshift=0.99cm] (d2) {\scriptsize$\bar{b}$};
					\vertex [above left=of a, xshift=0.55cm, yshift=-0.3cm] (a1) ;
					\vertex [below left=of a, xshift=0.55cm, yshift=0.3cm] (a2) ;
					\vertex [above left=of a1, xshift=0cm,yshift=-0.7cm] (a11) {\scriptsize$g$};
					\vertex [above right=of a1, xshift=0.15cm, yshift=-0.9cm] (b1) ;
					\vertex [above right=of b1, xshift=-0.7cm, yshift=-0.79cm] (d11) ;
					\vertex [below right=of b1, xshift=-0.1cm, yshift=0.99cm] (d12) {\scriptsize$b$} ;
					
					\vertex [above right=of d11, xshift=-1.16cm, yshift=-0.7cm] (d111) {\scriptsize$l^+$} ;
					\vertex [below right=of d11, xshift=-0.7cm, yshift=1.4cm] (d112) {\scriptsize$\nu_l$} ;
					
					\vertex [below left=of a2, xshift=0cm, yshift=0.7cm] (a21) {\scriptsize$g$};
					\vertex [below right=of a2, xshift=0.15cm, yshift=0.9cm] (b2) ;
					\vertex [above right=of b2, xshift=-0.1cm, yshift=-0.99cm] (d21) {\scriptsize$\bar{b}$} ;
					\vertex [below left=of b2, xshift=1.5cm, yshift=0.49cm] (d22)  ;
					
					\vertex [above right=of d22, xshift=-1.16cm, yshift=-1.87cm] (d121) {\scriptsize$l^-$} ;
					\vertex [below right=of d22, xshift=-0.7cm, yshift=1.1cm] (d122) {\scriptsize$\bar{\nu_l}$} ;
					
					\vertex [below right=of b2, xshift=-1.4cm, yshift=0.3cm]  (c) {\scriptsize$\gamma$} ;
					
					\diagram {
						(a) -- [scalar, line width = 0.02cm, edge label = \scriptsize$H$] (b);
						(d2) -- [fermion, line width = 0.02cm, arrow size = 0.03cm] (b) -- [fermion, line width = 0.02cm, arrow size = 0.03cm] (d1);
						(a2) -- [fermion, line width = 0.02cm, arrow size = 0.03cm, edge label = \scriptsize$t$] (a) -- [fermion, line width = 0.02cm, arrow size = 0.03cm, edge label = \scriptsize$\bar{t}$] (a1);
						(a1) -- [gluon, line width = 0.02cm] (a11);
						(a2) -- [gluon, line width = 0.02cm] (a21);
						(a1) -- [fermion, line width = 0.02cm, arrow size = 0.03cm, edge label = \scriptsize$t$] (b1);
						(b2) -- [fermion, line width = 0.02cm, arrow size = 0.03cm, edge label = \scriptsize$\bar{t}$] (a2);
						(b1) -- [boson, line width = 0.02cm, edge label = \scriptsize$W^+$] (d11);
						
						(d111) -- [fermion, line width = 0.02cm, arrow size = 0.03cm] (d11) -- [fermion, line width = 0.02cm, arrow size = 0.03cm] (d112);
						
						(d122) -- [fermion, line width = 0.02cm, arrow size = 0.03cm] (d22) -- [fermion, line width = 0.02cm, arrow size = 0.03cm] (d121);
						
						(b1) -- [fermion, line width = 0.02cm, arrow size = 0.03cm] (d12);
						(b2) -- [boson, line width = 0.02cm, edge label = \scriptsize$W^-$] (d22);
						(d21) -- [fermion, line width = 0.02cm, arrow size = 0.03cm] (b2);
						
					};
					\draw [photon] ($(a2)!0.6!(b2)$) -- (c);
				\end{feynman}
			\end{tikzpicture}
		}
	\end{minipage}
	\caption{Diagrams showing one leading-order contribution to the production of \tHa\, and \ttHa final states, with top quark decaying leptonically and $H \to b\bar{b}$.}
	\label{ttha}
\end{figure*}
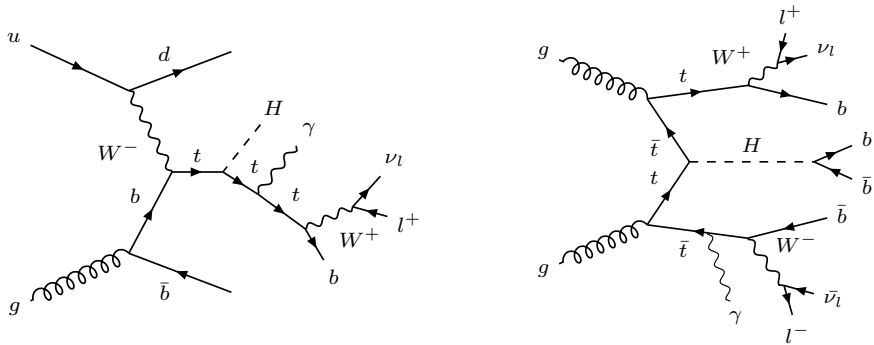

Compared to the other SM fermions, the top quark exhibits the strongest interaction with the Higgs boson, which is one of its noteworthy features. Directly measuring the Higgs boson’s decay into a top quark is not feasible, as the top quark’s mass exceeds that of the mass of Higgs boson. However, its coupling can be constrained through Higgs production processes that result in events where a Higgs boson is produced in association with a single top quark or a top-quark pair, as illustrated in the Figure \ref{ttha}. The associated production of a photon means the production of a cleaner signal and lesser background noise.

In this feasibility study, we explore the potential to search for the \tHa\, and \ttHa\, processes in final states with one or two leptons, a photon, and b-jets originating from Higgs boson decay, using simulated data from MadGraph \cite{madgraph}. In each channel, we reconstruct the \tHa\, and \ttHa\, system, accounting for the missing transverse energy (\MET) from neutrinos in leptonic top quark decays. The presence of the photon further aids in identifying the signal and distinguishing it from background processes.

The top quarks and Higgs boson must be reconstructed explicitly from the decay products of their final-state in order to directly study the top-quark Yukawa interaction. This condition is met by the associated production of a Higgs boson with a photon and either a single top-quark\,(\tHa) or a top-quark pair\,($t\bar{t}H\gamma$). The coupling between the top quark and the Higgs boson can be directly tested by measuring the production rate of \tHa\, and $t\bar{t}H\gamma$. Furthermore, a number of new physics scenarios \cite{heavytop2,heavytop3} suggest that heavy top-quark companions might exist and decay into Higgs boson and a top quark. An indirect indication of an unidentified phenomenon would be the observation of a significant discrepancy between the \tHa\, or $t\bar{t}H\gamma$ production rate and the SM prediction. This work describes the outcomes of the simulated production rate of \tHa\, and $t\bar{t}H\gamma$ in SM utilizing Delphes \cite{delphes} for the CMS detector effects.

\section{Simulation Samples}
\label{sec:Simulation_Samples}
Expected signal events and background processes are modelled with MC simulation. The signal processes \tHa\ and \ttHa, shown in Figure \ref{ttha}, as well as background processes listed in Tables \ref{tab:BG1} and \ref{tab:BG2}, are all generated with the matrix element generator, MadGraph5\_aMC@NLO\,\cite{madgraph}, PYTHIA\,8.3 \cite{pythia8.3} is integrated for hadronization and parton showering. For the \ttHa\ process, the cross section is computed up to next-to-leading order in QCD to account for higher order corrections. Samples produced using a leading order generator utilize the NNPDF23\_nlo\_as\_0119 Parton Distribution Function (PDF) set \cite{lhapdf6}, whereas the samples produced using NLO generators utilize the NNPDF23\_nlo\_as\_0118\_qed PDF set \cite{lhapdf6}. The detector response of CMS is simulated using DELPHES 3 \cite{delphes}. Both signal and background events are subjected to same selection criteria and undergo reconstruction using the same algorithms, as detailed in Section \ref{sec:Object_reconstruction_and_identification}.

The predicted event yields for the signal and background processes are listed in Tables \ref{tab:BG1} and \ref{tab:BG2}, together with their corresponding cross sections based on theoretical calculations and expected event counts. The events are calculated for an integrated luminosity of 350\,\text{fb$^{-1}$} at a centre-of-mass energy of 13.6\,TeV. The representative subprocesses contributing to  $pp \to t(\bar{t})\,H\,\gamma\,\bar{b}(b)\,q$ and $pp \to t\,\bar{t}\,H\,\gamma$ final states are listed in Table \ref{tab:subprocesses}, along with their respective cross sections.

\begin{table}[bht]
	\centering
	\caption{Signal and background processes with their respective cross sections and expected number of events for \tHa\,\,final states\,(L\,=\,350\,fb$^{-1}$, $\sqrt{s}$\,=\,13.6\,TeV).}
	\begin{tabular}{@{}>{\raggedright\arraybackslash}p{1.77cm}
			>{\centering\arraybackslash}p{3.47cm}
			>{\centering\arraybackslash}p{2.42cm}@{}}
		\hline
		\rule{0pt}{0pt} \\[-9pt]
		\multirow{2}{*}{Process} & Cross section\,[fb] & Expected Events \\
		& @ \footnotesize{13.6 TeV} & @ \footnotesize{350\,fb$^{-1}$} \\		
		\hline
		\rule{0pt}{0pt} \\[-9pt]
		$t\bar{b}\gamma$ & 16.91 & 5918 \\
		$t\bar{t}Z$ & 656.4 & $2.30\times10^{5}$ \\
		$WW$ & 68440 & $2.40\times10^{7}$ \\
		$t\bar{t}H$ & 393 & $1.38\times10^{5}$ \\
		$t\bar{t}bb$ & 15450 & $5.41\times10^{6}$ \\
		$WW\gamma$ & 282.3 & $9.88\times10^{4}$ \\
		$t\bar{t}jj$ & 481400 & $1.68\times10^{8}$ \\
		$t\bar{t}jj\gamma$ & 2312 & $8.09\times10^{5}$ \\
		$t\bar{t}Z\gamma$ & 5.1 & 1793 \\
		$t\bar{t}H\gamma$ & 3.082 & 1079 \\
		$t\bar{t}bb\gamma$ & 72.2 & $2.53\times10^{4}$ \\
		$tH\bar{b}j$ & 41.91 & $1.47\times10^{4}$ \\
		$\bar{t}bj$ & 107600 & $3.77\times10^{7}$ \\
		$Zj\gamma$ & 33540 & $1.17\times10^{7}$ \\
		$ZZ\gamma$ & 43.95 & $1.54\times10^{4}$ \\
		\hline
		\rule{0pt}{0pt} \\[-8pt]
		\text{Total bkg} &&  $2.48\times10^{8}$ \\
		\hline
		\rule{0pt}{0pt} \\[-8pt]
		\text{Expected Signal} & 0.2767 & 96 \\
		\hline
	\end{tabular}
	\label{tab:BG1}
\end{table}

The full list of input variables used for BDT training is provided in Table \ref{tab:bdt_appendix}. Each variable is normalized to the total integrated luminosity of 350\,fb$^{-1}$. These variables include kinematic and event-level properties, such as the transverse momentum sum (\(\Sigma p_T\)) of photons, leptons, b-jets, jets, light-jets, \MET, as well as counts of leptons, jets and photons. Specific features, such as the $m_{jj}$ and $\Delta\eta_{jj}$ are included to improve signal-background separation. The variable $m_{jj}$ is defined as the invariant mass of the jet pair whose value is closest to the Higgs boson mass among all possible jet pairs in the event, where $\Delta\eta_{jj}$ refers to the pseudorapidity difference between the jet pairs.

\begin{figure}[hbt]
	\centering
	\begin{minipage}{0.48\linewidth}
		\centering
		\vspace{0.08\textwidth}
		\includegraphics[width=0.71\linewidth]{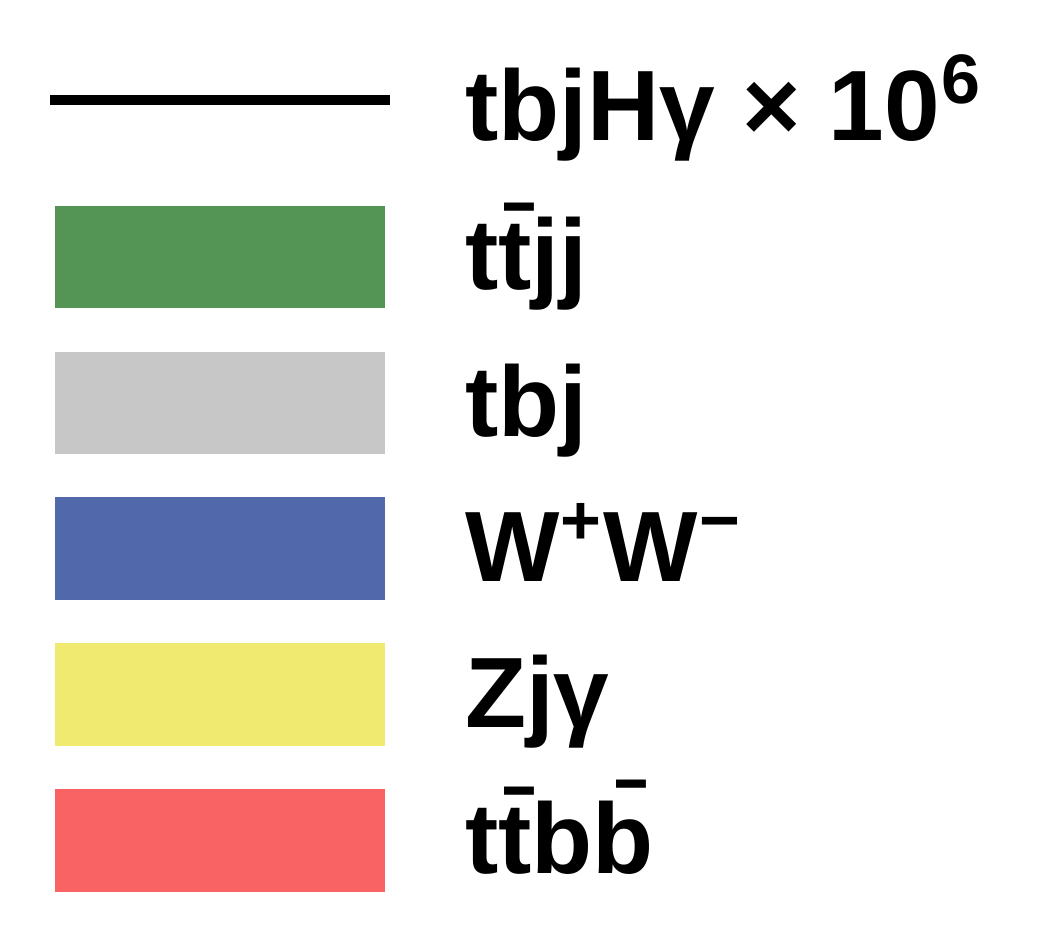} \\  
		\vspace{0.08\textwidth}
		\includegraphics[width=1.07\linewidth]{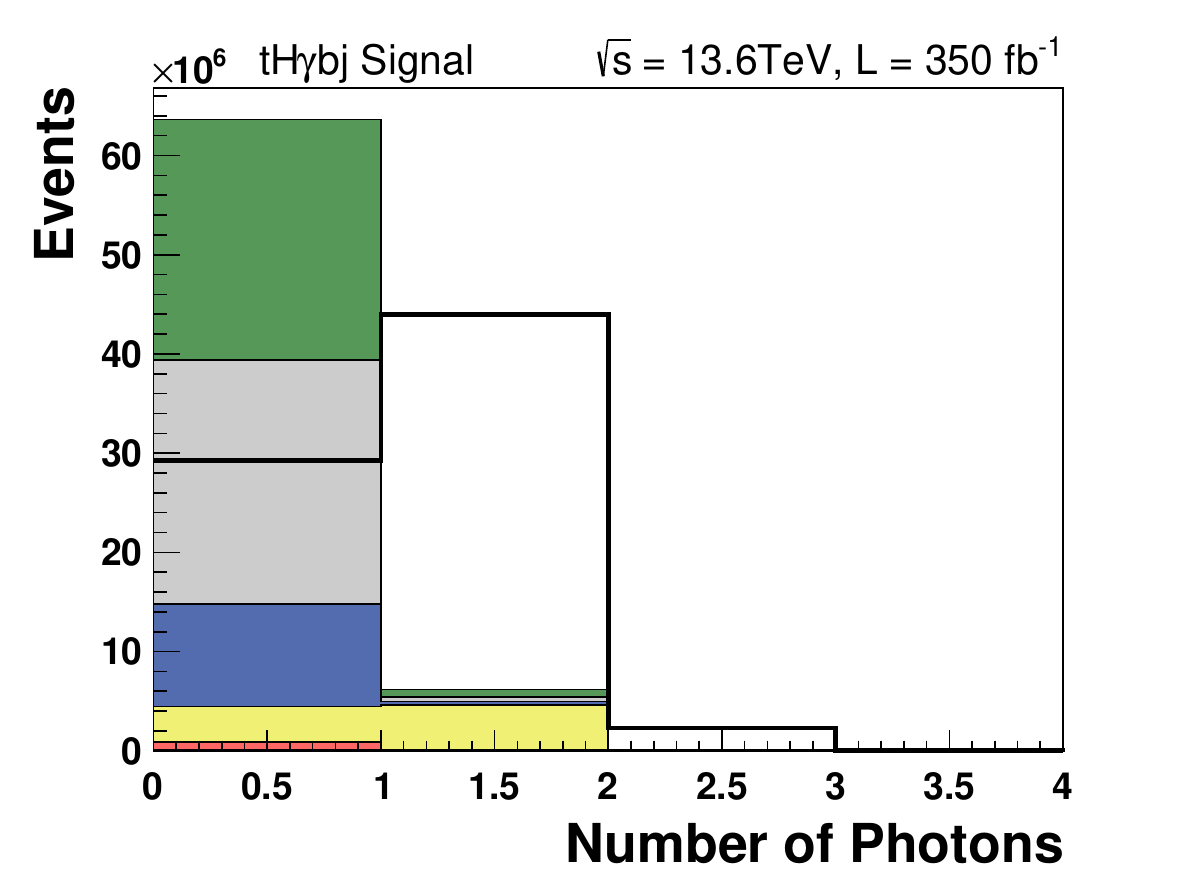} 
	\end{minipage}
	\begin{minipage}{0.49\linewidth}
		\centering 
		\includegraphics[width=1.07\linewidth]{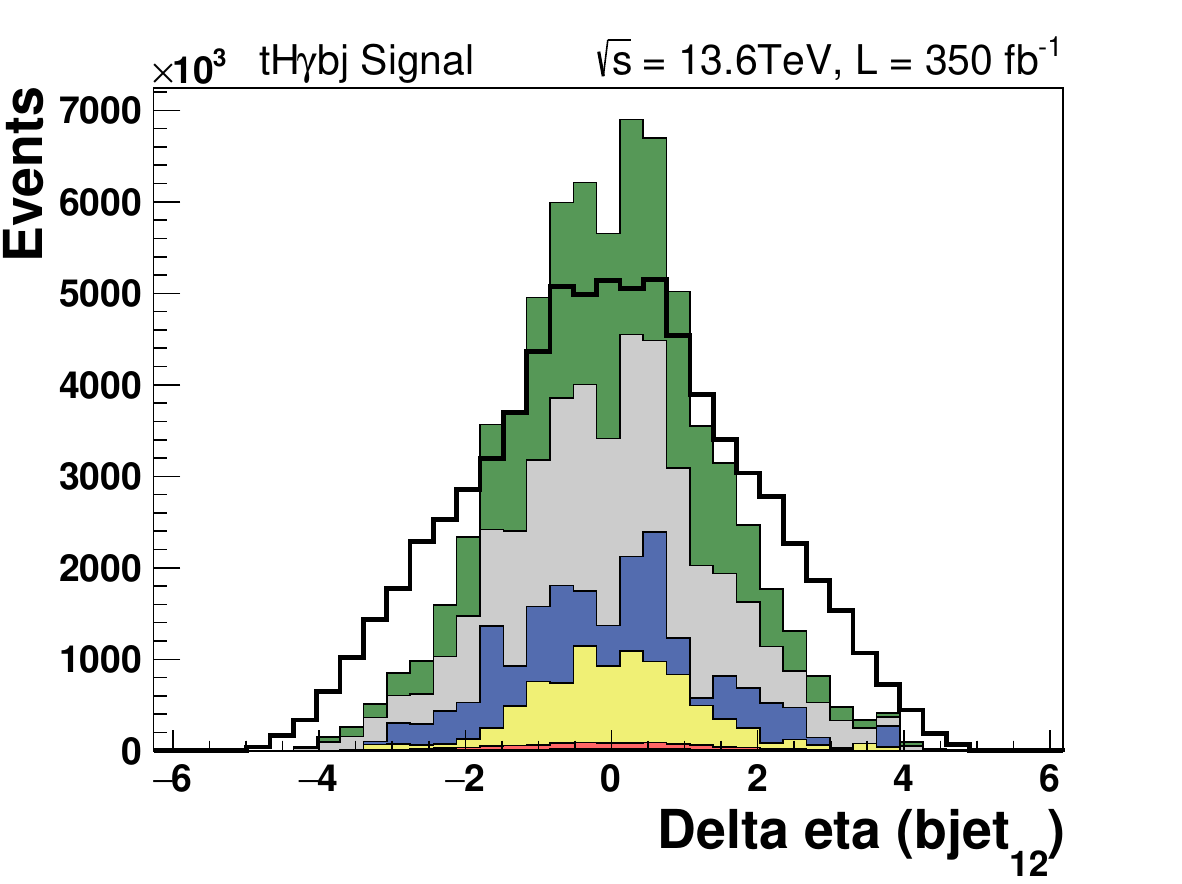}\\
		\includegraphics[width=1.07\linewidth]{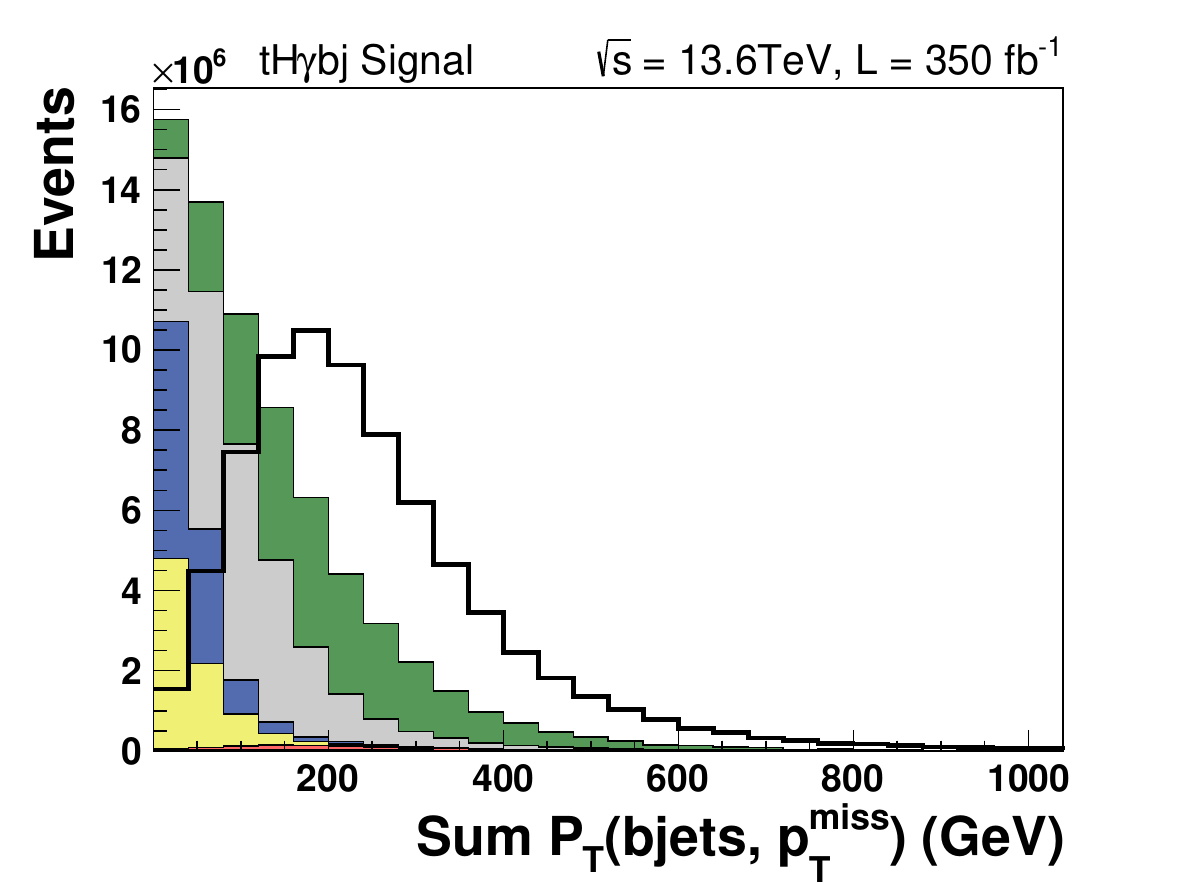} 
	\end{minipage}
	
	\caption{Comparison of $\Delta\eta_{b_{12}}$, photon multiplicity, and Sum of $p_T$ of bjets and \MET\ for signal and background events. These are the best performing variables for \tHa\ channel.}
	\label{fig:IV_tha}
\end{figure}

A selection of the most discriminating variables is visualized in Figures \ref{fig:IV_tha} and \ref{fig:IV_ttha}. These plots reveal differences in the distribution of input features between signal and background events, driven by the event topologies of the \tHa\, and \ttHa\, processes.

\begin{figure}[hbt]
	\centering
	\begin{minipage}{0.482\linewidth}
		\centering
		\vspace{3em}
		\includegraphics[width=\linewidth]{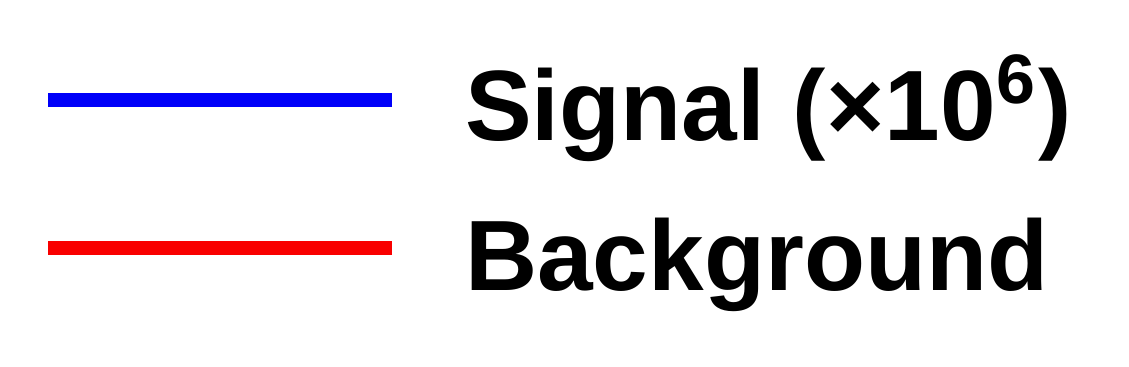} \\ 
		\vspace{2.7em}
		\includegraphics[width=\linewidth]{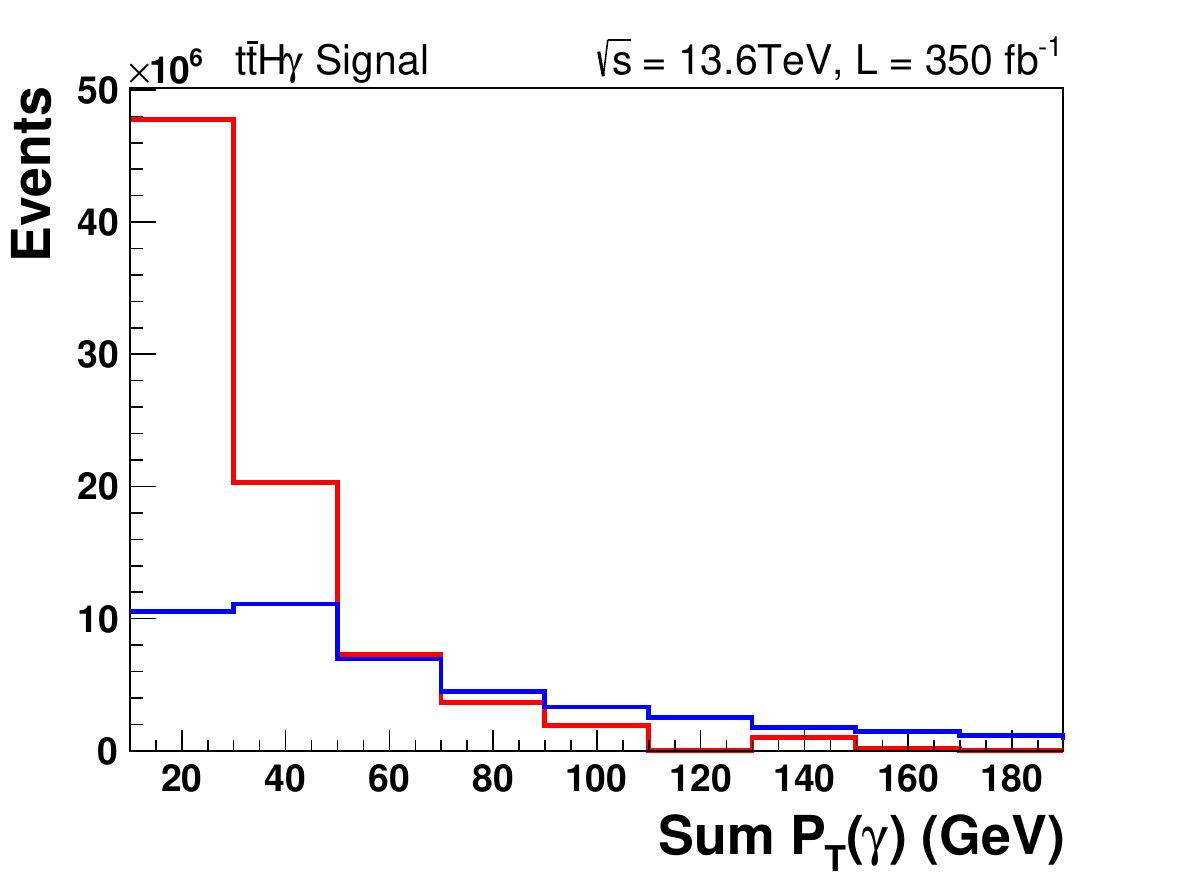} 
	\end{minipage}
	\begin{minipage}{0.49\linewidth}
		\centering 
		\includegraphics[width=\linewidth]{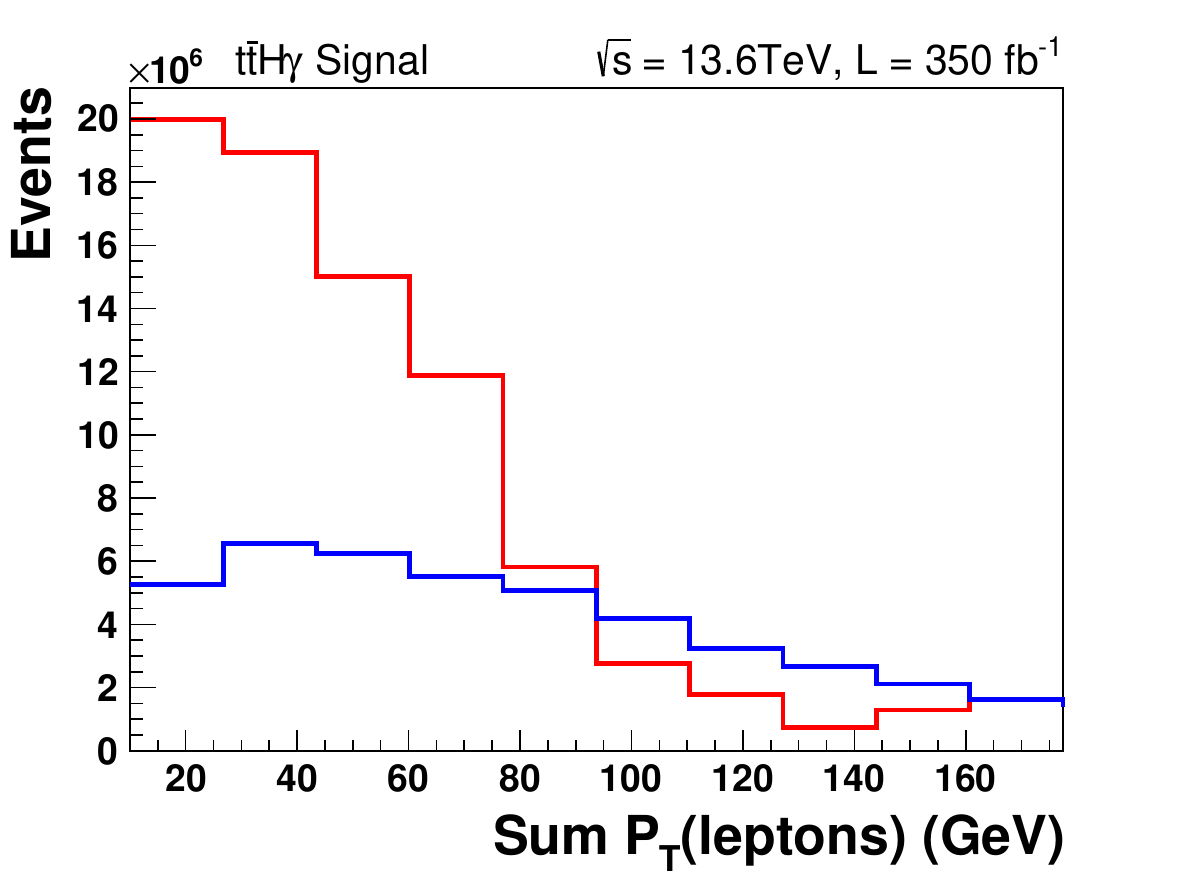}\\
		\includegraphics[width=\linewidth]{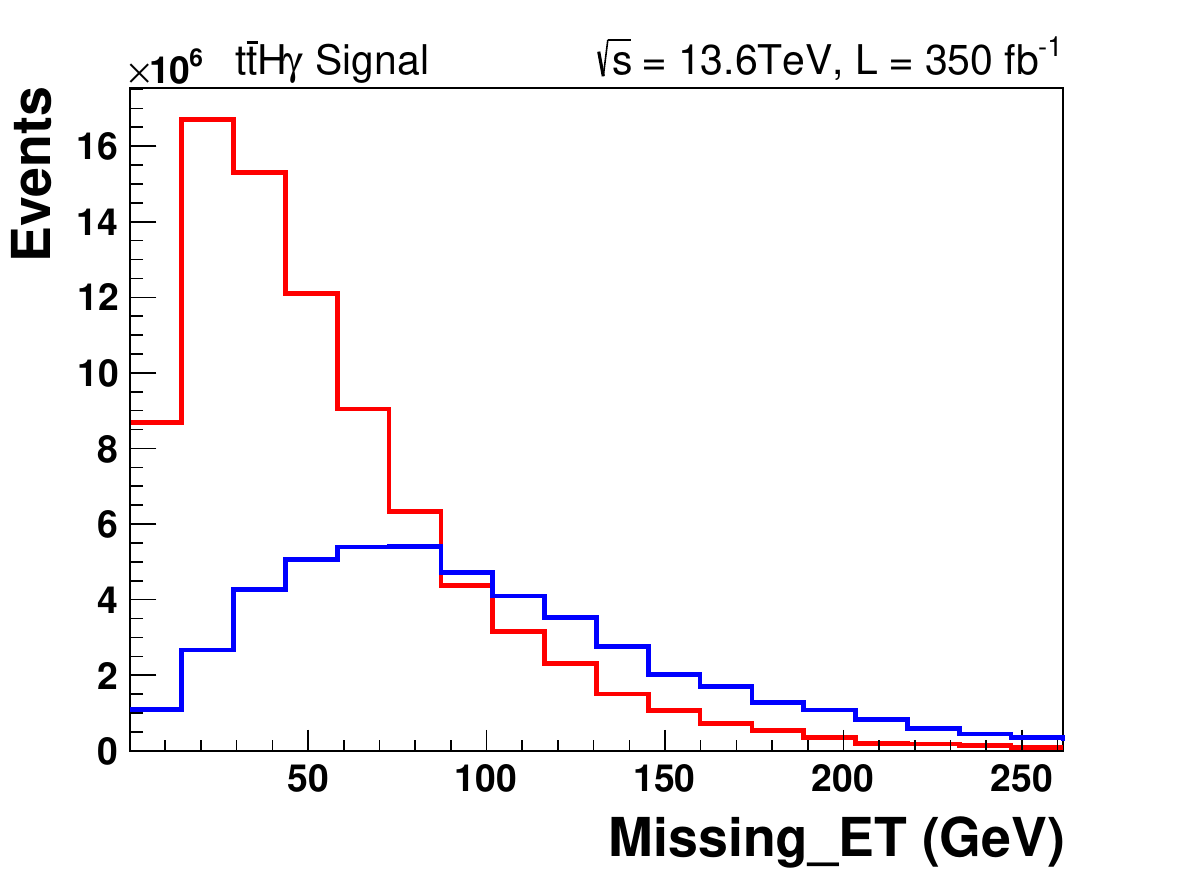} 
	\end{minipage}
	
	\caption{Distributions of the $\Sigma p_T$ of leptons, $\Sigma p_T$ of photons and \MET\ for \ttHa\ signal events compared to dominant backgrounds. The blue line represents the signal, while the red line represents the sum of all the backgrounds, normalized according to their respective cross sections.}
	\label{fig:hists_ttha}
\end{figure}

One key variable is the number of photons, which tends to be higher in signal events. While both signal and background events may contain photons from initial or final-state radiation or from misidentified jets, the signal processes explicitly produce a photon, resulting in higher photon multiplicity, as shown in photon multiplicity plots in Figures \ref{fig:IV_tha} and \ref{fig:IV_ttha}. Similarly, the number of leptons serves as an important discriminator. In signal events, the top quarks decay leptonically, leading to a higher lepton multiplicity compared to most background processes, which typically contain fewer leptons.

\begin{figure}[hbt]
	\centering
	\includegraphics[width=0.49\linewidth]{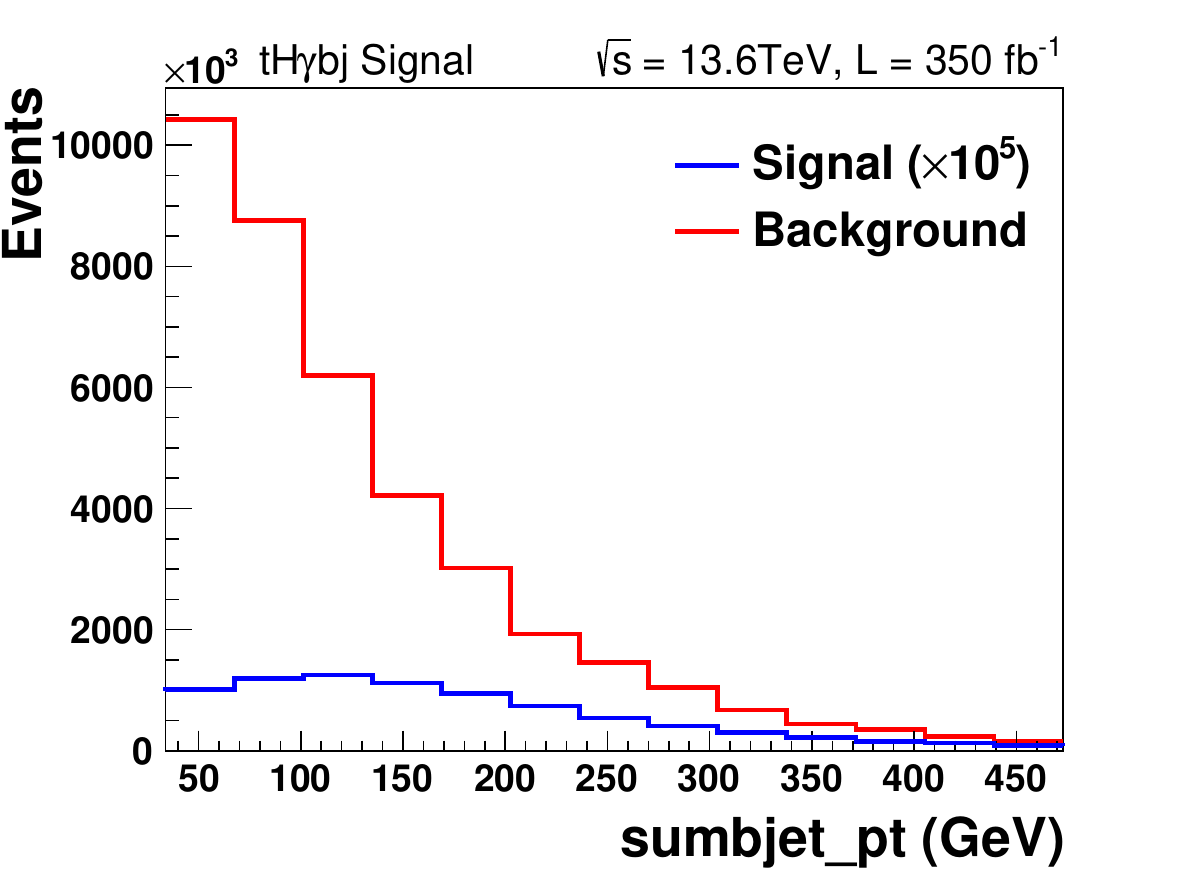}
	\includegraphics[width=0.49\linewidth]{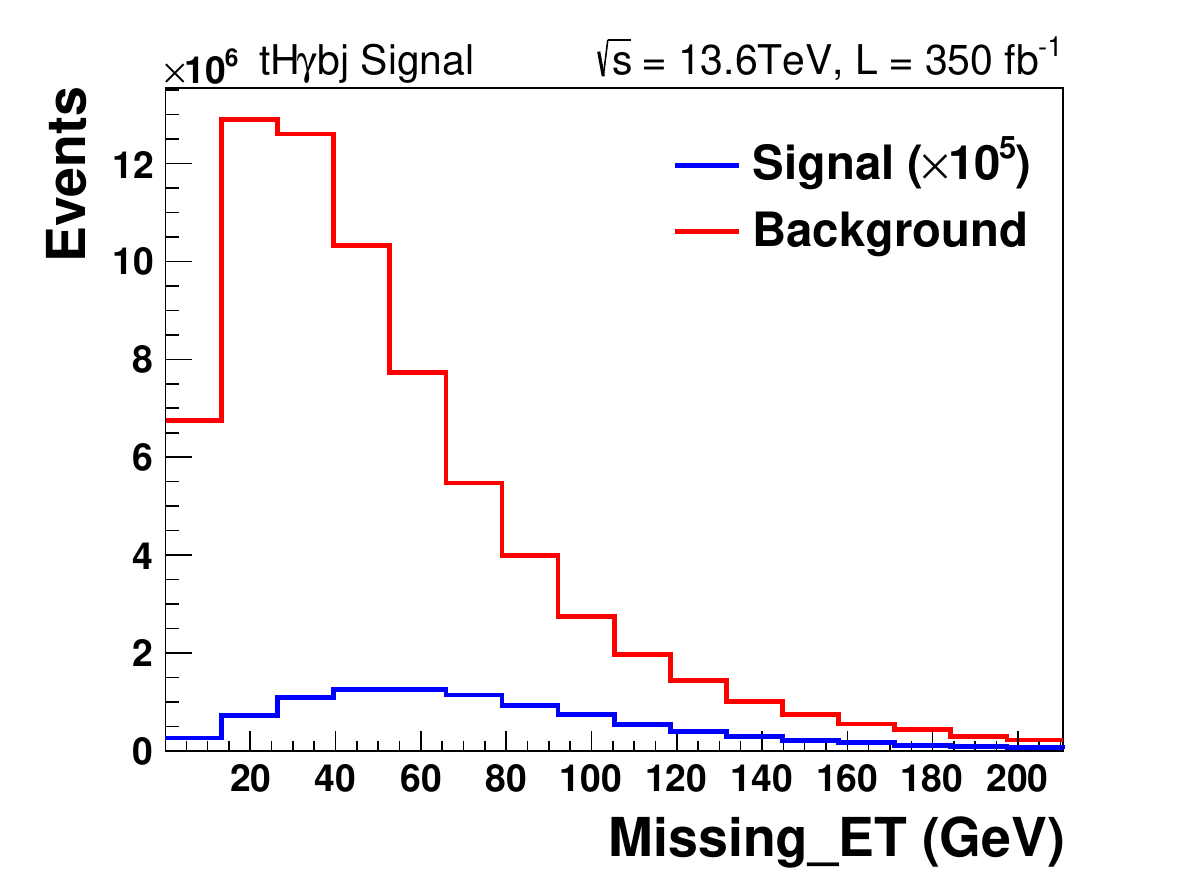}
	\caption{Distributions of the $\Sigma p_T$ of b-jets (left) and \MET\ (right) for \tHa\ signal events compared to dominant backgrounds. The blue line represents the signal, while the red line represents the sum of all the backgrounds, normalized according to their respective cross sections.}
	\label{fig:hists_tha}
\end{figure}
\vspace{-0.4cm}
\begin{table}[bht]
	\centering
	\caption{Signal and background processes with their respective cross sections and expected number of events for \ttHa\,\,final states\,(L\,=\,350\,$\text{fb$^{-1}$}$, $\sqrt{s}$\,=\,13.6\,TeV).}
	\begin{tabular}{@{}>{\raggedright\arraybackslash}p{1.77cm}
			>{\centering\arraybackslash}p{3.47cm}
			>{\centering\arraybackslash}p{2.42cm}@{}}
		\hline
		\rule{0pt}{0pt} \\[-9pt]
		\multirow{2}{*}{Process} & Cross section\,[fb] & Expected Events \\
		& @ \footnotesize{13.6 TeV} & @ \footnotesize{350\,fb$^{-1}$} \\
		\hline
		\rule{0pt}{0pt} \\[-9pt]
		\text{Diboson} & 78379 & $2.74\times10^{7}$ \\
		\text{Triboson} & 207 & $7.25\times10^{4}$ \\
		\text{Diboson + $\gamma$} & 422 & $1.48\times10^{5}$ \\
		\text{Triboson + $\gamma$} & 2.81 & 986 \\
		$t\bar{t}H$ & 393 & $1.38\times10^{5}$ \\
		$t\bar{t}$ & 511000 & $1.79\times10^{8}$ \\
		{$t\bar{t}\gamma$} & 2273 & $7.96\times10^{5}$ \\
		$t\bar{t}b\bar{b}$ & 15450 & $5.41\times10^{6}$ \\
		$t\bar{t}b\bar{b}\gamma$ & 72.2 & $2.53\times10^{4}$ \\
		$t\bar{t}jj$ & 481400 & $1.68\times10^{8}$ \\
		{$t\bar{t}jj\gamma$} & 2312 & $8.09\times10^{5}$ \\
		\text{$b\bar{b}WW$} & 597900 & $2.09\times10^{8}$ \\
		$b\bar{b}WW\gamma$ & 5865 & $2.05\times10^{6}$ \\
		$tbWH$ & 903 & $3.16\times10^{5}$ \\
		$tbWH\gamma$ & 9.26 & 3241 \\
		$t\bar{t}Z\gamma$ & 5.1 & 1793 \\
		\hline
		\rule{0pt}{0pt} \\[-8pt]
		\text{Total bkg} &&  $6.03\times10^{8}$ \\
		\hline
		\rule{0pt}{0pt} \\[-8pt]
		\text{Expected Signal} & 0.139 & 48 \\
		\hline
		
	\end{tabular}
	\label{tab:BG2}
\end{table}

Signal events are characterized by higher values of $\Sigma p_T$ for objects such as photons, leptons, and b-jets, primarily due to their greater multiplicities compared to background events. Additionally, the \MET\ tends to be larger in signal events because of neutrinos produced in leptonic top quark decays. These features are illustrated in Figures \ref{fig:hists_tha} and \ref{fig:hists_ttha} for the \tHa\ and \ttHa\ processes respectively. While each variable provides some discrimination power individually, combining them to create a derived variable significantly improves overall separation power. Two of these derived variables, $\Sigma p_T$ of b-jets, and \MET, which performed the best for \tHa\ dataset, and $\Sigma p_T$ of photons, leptons, and \MET, for \ttHa\ dataset, are shown in Figures \ref{fig:IV_tha} and \ref{fig:IV_ttha}, respectively.

The variable $\Delta\eta_{b_{12}}$, shown in Figure \ref{fig:IV_tha}, representing the pseudorapidity difference between the two leading b-jets, is used to exploit the fact that in signal events, b-jets often originate from different sources, one from a top decay and the other from a Higgs decay, resulting in a broader angular separation. Background events, typically produce more central and collimated b-jets, leading to smaller $\Delta\eta$ values.
\vspace{-0.20cm}
\section{Object reconstruction and identification}
\vspace{-0.10cm}
\label{sec:Object_reconstruction_and_identification}
For the recognition and reconstruction of the \tHa\, and \ttHa\, events, various particle types are identified and reconstructed using Delphes, which simulates a fast, parametrized detector response of the CMS detector. Delphes mimics the CMS performance using simplified detector geometries and smearing functions, and applies reconstruction algorithms approximating those based on the Particle Flow\,(PF) technique at the CMS \cite{PF_reco,PF2,PF1}. These particles include neutral and charged hadrons clustered into jets, muons, electrons, photons and \MET\ attributed to neutrinos.

\begin{table*}[t]
	\centering
	\renewcommand{\arraystretch}{1.2}
	\caption{Input variables and their feature importance rankings for distinguishing \tHa\, and \ttHa\, signals from backgrounds using XGBoost and TMVA BDTs. A dash (--) indicates the variable was not used or not ranked.}
	\label{tab:bdt_appendix}
	\begin{tabular}{@{}
			>{\raggedright\arraybackslash}p{7.2cm}  
			>{\centering\arraybackslash}p{2cm}      
			>{\centering\arraybackslash}p{2.4cm}    
			>{\centering\arraybackslash}p{2.1cm}    
			>{\centering\arraybackslash}p{2.5cm}@{}}
		\hline
		\multirow{2}{*}{\centering\arraybackslash BDT Inputs} 
		& \multicolumn{2}{c}{\tHa} 
		& \multicolumn{2}{c}{\ttHa} \\
		& XGBoost & TMVA & XGBoost & TMVA \\
		\hline
		$\Sigma p_T$ of photons, leptons, \MET\,and jets            & 1  & -- & -- & -- \\
		$\Sigma p_T$ of photons, leptons and \MET                  & -- & -- & 1  & -- \\
		$\Sigma p_T$ of bjets and \MET                             & 2  & 8  & -- & -- \\
		$\Sigma p_T$ of bjets                                      & -- & -- & 3  & -- \\
		$\Sigma p_T$ of Jets                                       & 8  & 11 & -- & -- \\
		$\Delta\eta_{b_{12}}$                                      & 7  & 1  & -- & -- \\
		$\Delta\eta_{b_{23}}$                                      & -- & 4  & -- & -- \\
		$\Delta\eta_{j_{12}}$                                      & 11 & 2  & -- & -- \\
		$\Delta\eta_{j_{23}}$                                      & 12 & 5  & -- & -- \\
		Number of leptons                                          & 3  & -- & 2  & 6  \\
		Number of Photons                                          & 15 & 3  & 4  & 1  \\
		Number of light-jets                                       & 16 & 7  & -- & 5  \\
		Number of bjets                                            & 10 & 10 & -- & 4  \\
		Number of Jets                                             & 14 & 9  & 12 & -- \\
		Number of Muons                                            & -- & 12 & -- & -- \\
		Number of Electrons                                        & -- & 16 & -- & -- \\
		Ratio of \MET\,and Scalar HT                               & 6  & -- & 11 & -- \\
		Missing Transverse Energy (\MET)                           & 9  & 6  & 13 & -- \\
		Scalar HT                                                  & 5  & 13 & -- & -- \\
		$p_T$ of 1st subleading Jet                                & 4  & 14 & -- & -- \\
		$p_T$ of Leading Jet                                       & 13 & 15 & 6  & -- \\
		$\eta$ of bjets                                            & -- & -- & 5  & -- \\
		Jet pair with mass third closest to 125\,GeV              & -- & -- & 7  & -- \\
		$\Sigma p_T$ of photons, ljets and \MET                   & -- & -- & 8  & -- \\
		$\Sigma p_T$ of photons, Scalar HT and jets              & -- & -- & 9  & -- \\
		$\Sigma p_T$ of photons, \MET\,and jets                    & -- & -- & 10 & -- \\
		$\Sigma p_T$ of photons, Scalar HT, leptons and \MET      & -- & -- & -- & 2  \\
		$\Sigma p_T$ of photons, Scalar HT, ljets and leptons     & -- & -- & -- & 3  \\
		\hline
	\end{tabular}
\end{table*}

Delphes mimics CMS by reconstructing photons from clustering energy deposits in the electromagnetic calorimeter (ECAL). A photon candidate is selected if it has a transverse momentum $p_T > 10\,\text{GeV}$ and lies within the pseudorapidity range $\lvert\eta\rvert < 2.4$. Photon identification includes isolation criteria based on surrounding PF candidates, simulating the isolation requirements used in CMS \cite{delphes,PF1,diphoton}.

Electrons are reconstructed by matching ECAL energy deposits to inner tracking information. Delphes applies smearing to the electron energy and uses efficiency maps as a function of $p_T$ and $\eta$ to emulate the reconstruction performance. Electrons are accepted if $p_T > 10\,\text{GeV}$ and $\lvert\eta\rvert < 2.4$. Delphes parametrizes the overall electron momentum resolution and efficiency to match CMS performance \cite{delphes,e_reco1,e_gamma_reco1,e_gamma_reco2}.

\begin{table}[bht]
	\centering
	\caption{Comparison of Test AUC values for TMVA BDT and XGBoost BDTG on \tHa\, and \ttHa\, datasets.}
	\begin{tabular}{|c|c|c|}
		\hline
		\rule{0pt}{2.5ex}
		Classifier & \tHa\ AUC (\%) & \ttHa\ AUC (\%) \\
		\hline
		\rule{0pt}{2.5ex}
		TMVA BDT          & 92 & 97 \\
		XGBoost BDTG      & 99 & 99 \\
		\hline
	\end{tabular}
	\label{tab:TestAUC_Comparison}
\end{table}

Muons are reconstructed by combining information from the muon system and the tracker. Delphes uses smearing functions for momentum and applies a reconstruction efficiency based on $\eta$ and $p_T$. Muon candidates are selected with $p_T > 10\,\text{GeV}$ and $\lvert\eta\rvert < 2.4$, matching the typical acceptance of the CMS muon system. The simulation assumes matching between inner and outer tracks without explicitly modelling the inside-out or outside-in track building strategy\,\cite{delphes,PF_muon}.

Jet reconstruction in Delphes is accomplished by parametrizing detector efficiency and smearing four-momentum information of particles. The anti-$k_T$ algorithm is utilized to reconstruct jets having a radius parameter R\,=\,0.4, within the pseudorapidity range $\vert\eta\rvert < 2.4$ and for transverse momenta greater than 20\,GeV. Jet energy and direction are smeared using detector resolution functions. Reconstruction efficiency reduces considerably outside the tracker's geometrical acceptance, where the calorimeter coverage is poor, or below the defined transverse momentum threshold \cite{delphes}.

B-jet identification in Delphes is achieved by implementing b-tagging techniques, which use particular algorithms to discriminate b-jets from light-flavoured and gluon-initiated jets. The simulation models b-tagging using parametrized efficiency and mistag rates based on pseudorapidity $\eta$ and jet $p_T$. These parametrizations account for the distinct kinematic and spatial properties of jets containing b-hadrons, such as the presence of a secondary vertex, displaced tracks, and higher masses \cite{delphes}.

\begin{table}[bht]
	\centering
	\caption{Cross sections for the individual subprocesses contributing to the production of $p,p \to t(\bar{t}),H,\gamma,\bar{b}(b),q$ and $p,p \to t,\bar{t},H,\gamma$ final states, along with the total cross section for these final state. The values correspond to decay-level cross sections (cross section $\times$ branching ratio).}
	\begin{tabular}{|c|l|c|}
		\hline
		\rule{0pt}{2.5ex}
		\tHa & subprocesses & Cross section [fb] \\
		\hline
		\rule{0pt}{2.5ex}
		1 & g, q $\to$ t, H, $\gamma$, $\bar{b}$, q & 0.183 \\
		2 & g, q $\to$ $\bar{t}$, H, $\gamma$, b, q & 0.094 \\
		\hline
		\rule{0pt}{2.5ex}
		& Total cross section                          & 0.277  \\
		\hline
		\rule{0pt}{2.5ex}
		\ttHa & subprocesses & Cross section [fb] \\
		\hline
		\rule{0pt}{2.5ex}
		1 & g, g $\to$ t, $\bar{t}$, H, $\gamma$ & 0.067 \\
		2 & u, $\bar{u}$ $\to$ t, $\bar{t}$, H, $\gamma$ & 0.029 \\
		3 & $\bar{u}$, u $\to$ t, $\bar{t}$, H, $\gamma$ & 0.029 \\
		4 & d, $\bar{d}$ $\to$ t, $\bar{t}$, H, $\gamma$ & 0.007 \\
		5 & $\bar{d}$, d $\to$ t, $\bar{t}$, H, $\gamma$ & 0.007 \\
		\hline
		\rule{0pt}{2.5ex}
		& Total cross section                          & 0.139  \\
		\hline
	\end{tabular}
	\label{tab:subprocesses}
\end{table}

In high-energy hadron collisions, the initial-state partons possess negligible transverse momentum. As a result, any observed imbalance in the total transverse momentum of the final-state particles is indicative of undetected components, such as neutrinos, and is interpreted as missing transverse energy (\MET). The vector \MET\,is defined as:
\begin{equation}
	\vec{\mspace{2mu}\mathrm{p}}_{\mathrm{T}}^{\mathrm{miss}} = - \sum_i \vec{\mspace{2mu}\mathrm{p}}_{\mathrm{T}}^{\mathrm{(i)}}
	\label{missing_ET}
\end{equation}
Delphes uses different reconstruction methods to compute \MET. Calorimeter based \MET\,is reconstructed using energy deposits in calorimeter towers only. The Particle-Flow (PF) algorithm combines information from various detector subsystems, such as calorimeters and tracking chambers, to reconstruct individual particles and accurately determine their properties, including energy and momentum. The PF \MET\, is reconstructed using information from PF towers and PF tracks. Reconstruction of \MET\,is crucial for identifying processes with undetected particles \cite{delphes}.

\begin{figure}[hbt]
	\centering
	\includegraphics[width=0.47\linewidth]{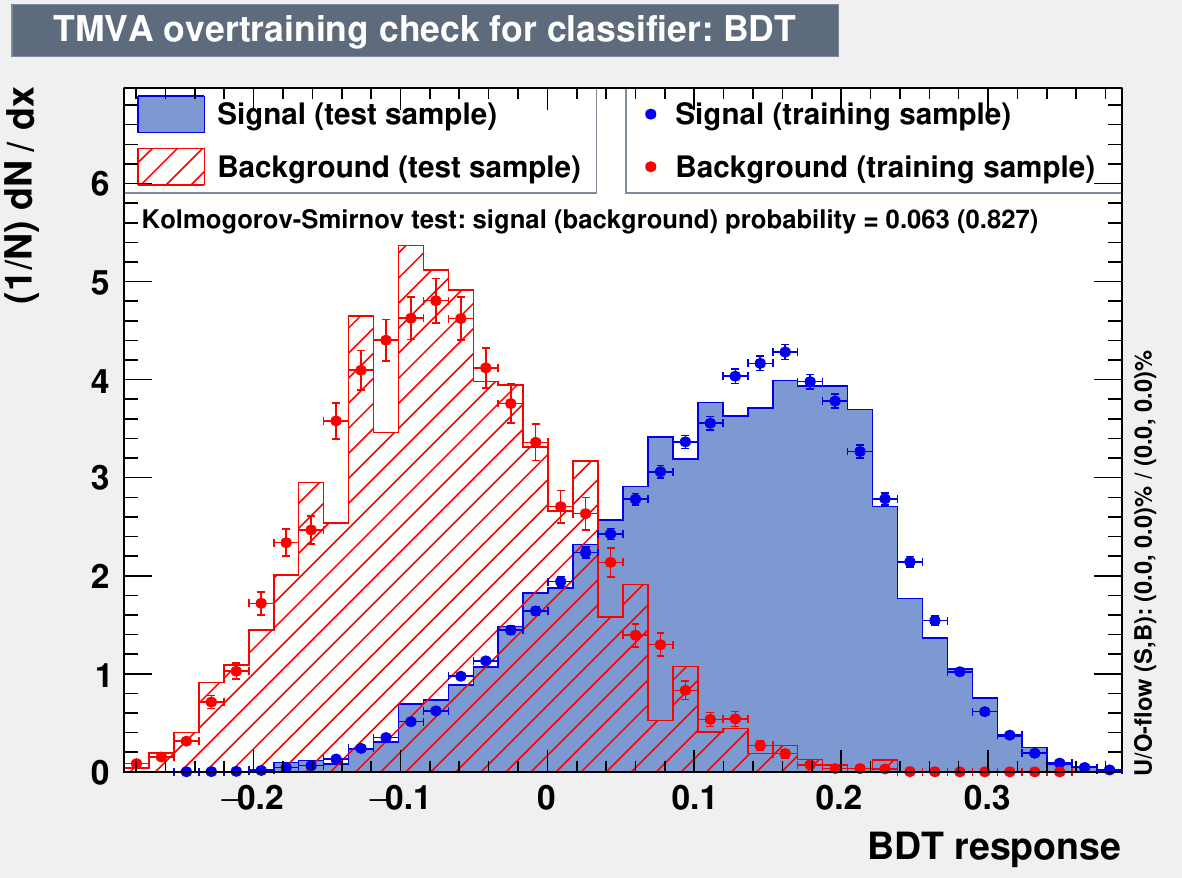}
	\hspace{0.01\linewidth} 
	\includegraphics[width=0.47\linewidth]{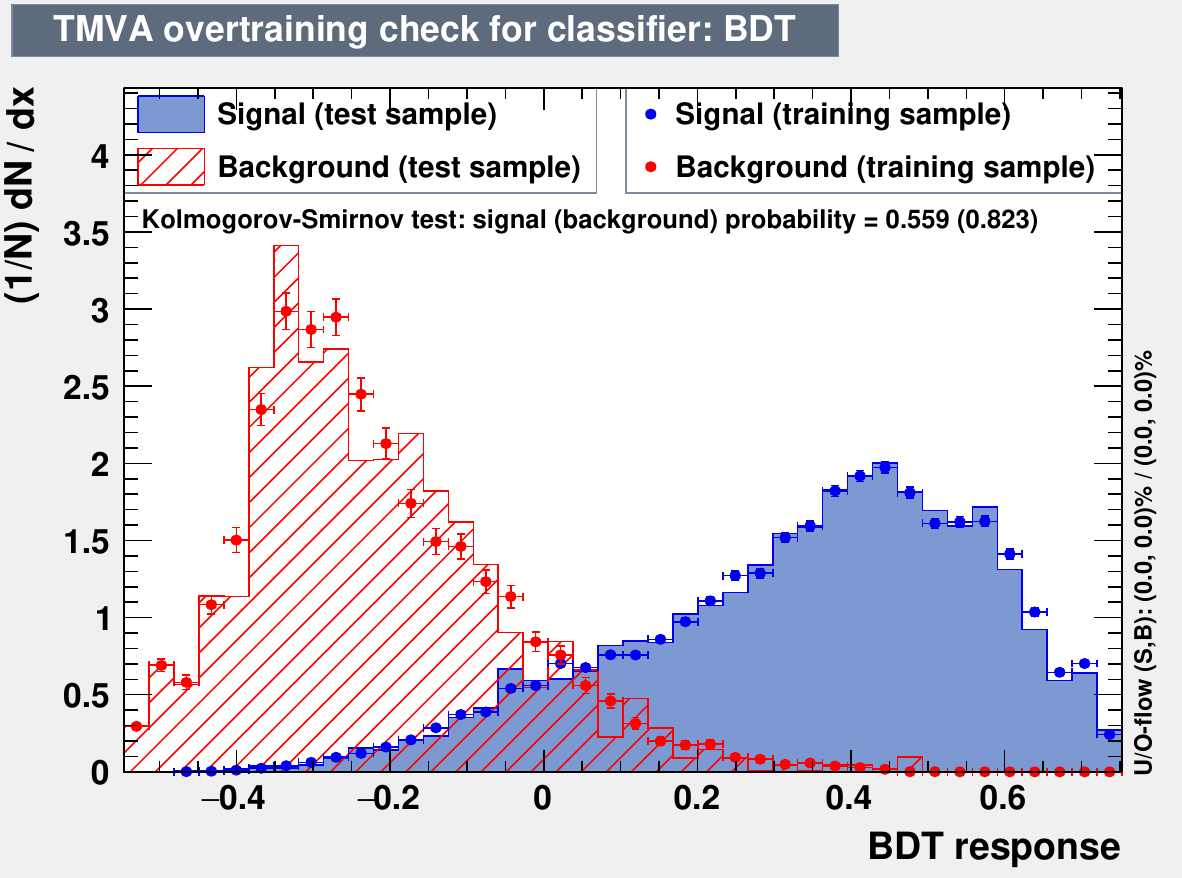}
	\caption{Overtraining Evaluation for TMVA BDT Classifier for \tHa\,process\,(left), \ttHa\,process\,(right).}
	\label{fig:OT_TMVA_tha_ttha}
\end{figure}

\section{Signal Extraction}
\begin{figure}[bht]
	\centering
	\begin{minipage}{0.48\linewidth}
		\centering
		\vspace{0.07\linewidth}
		\includegraphics[width=0.80\linewidth]{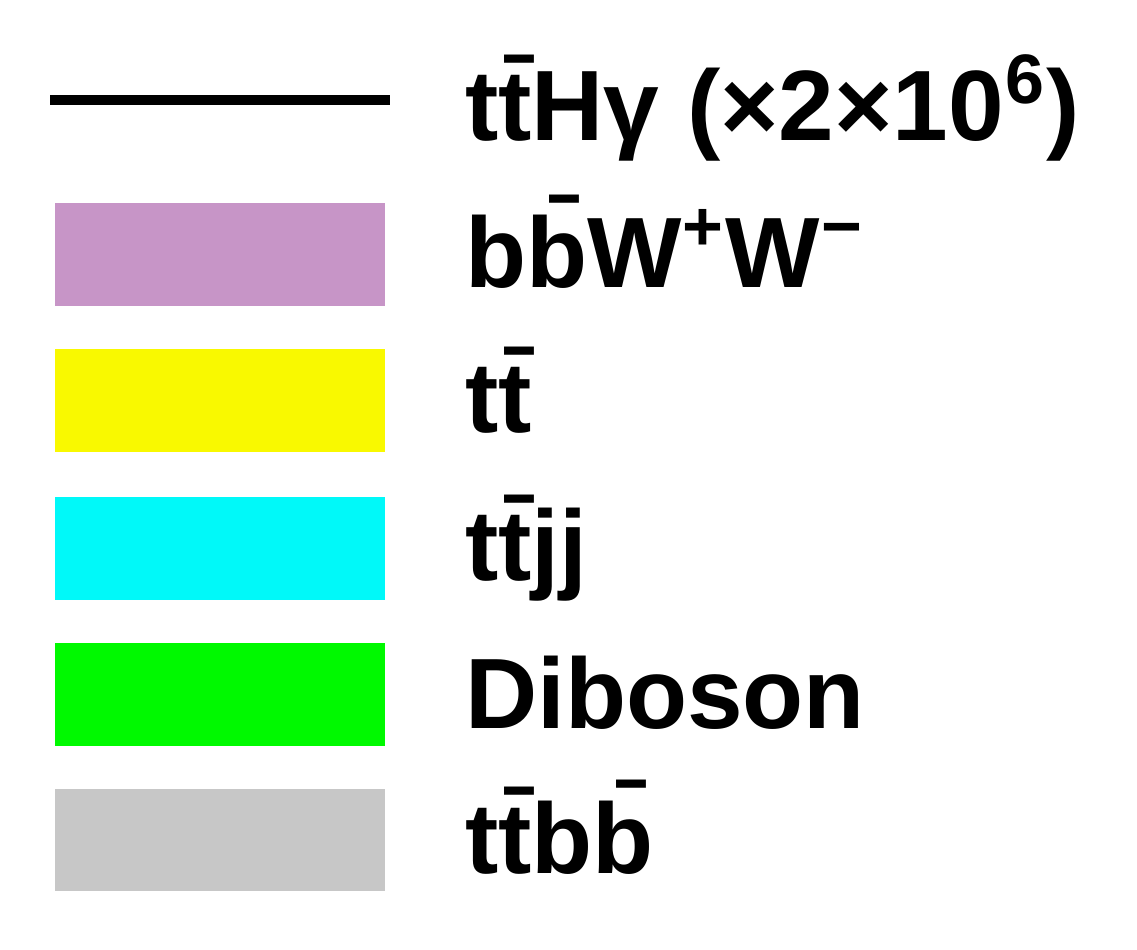} \\  
		\vspace{0.06\linewidth}
		\includegraphics[width=1.05\linewidth]{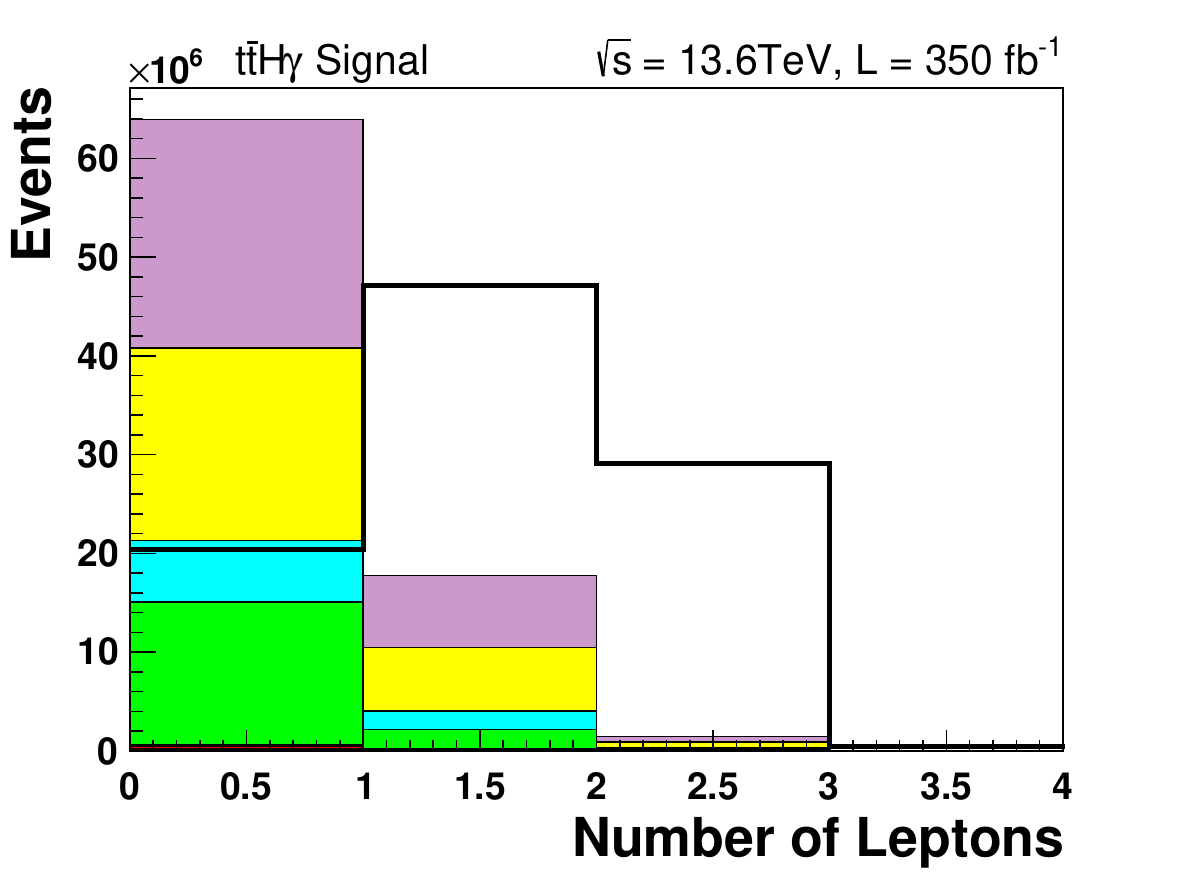} 
	\end{minipage}
	\begin{minipage}{0.49\linewidth}
		\centering
		\includegraphics[width=1.05\linewidth]{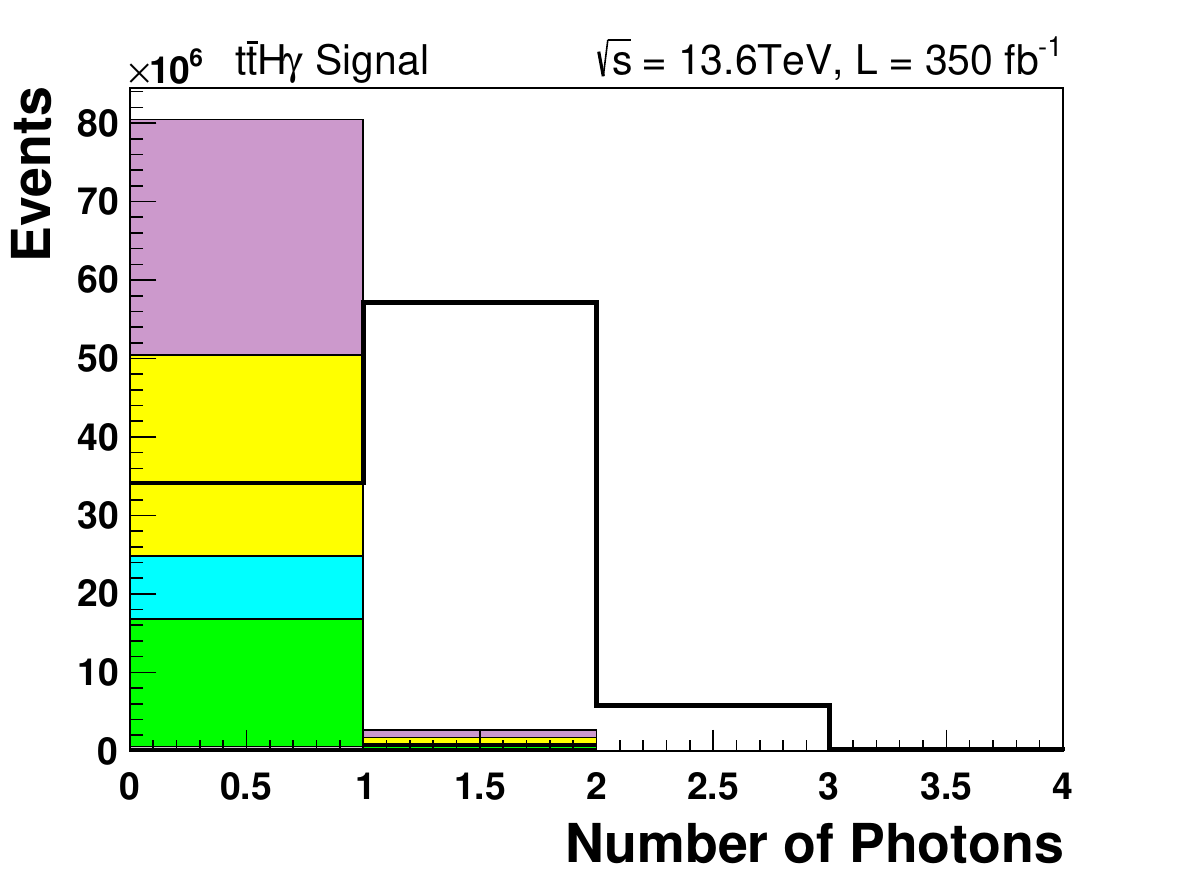} \\
		\includegraphics[width=1.05\linewidth]{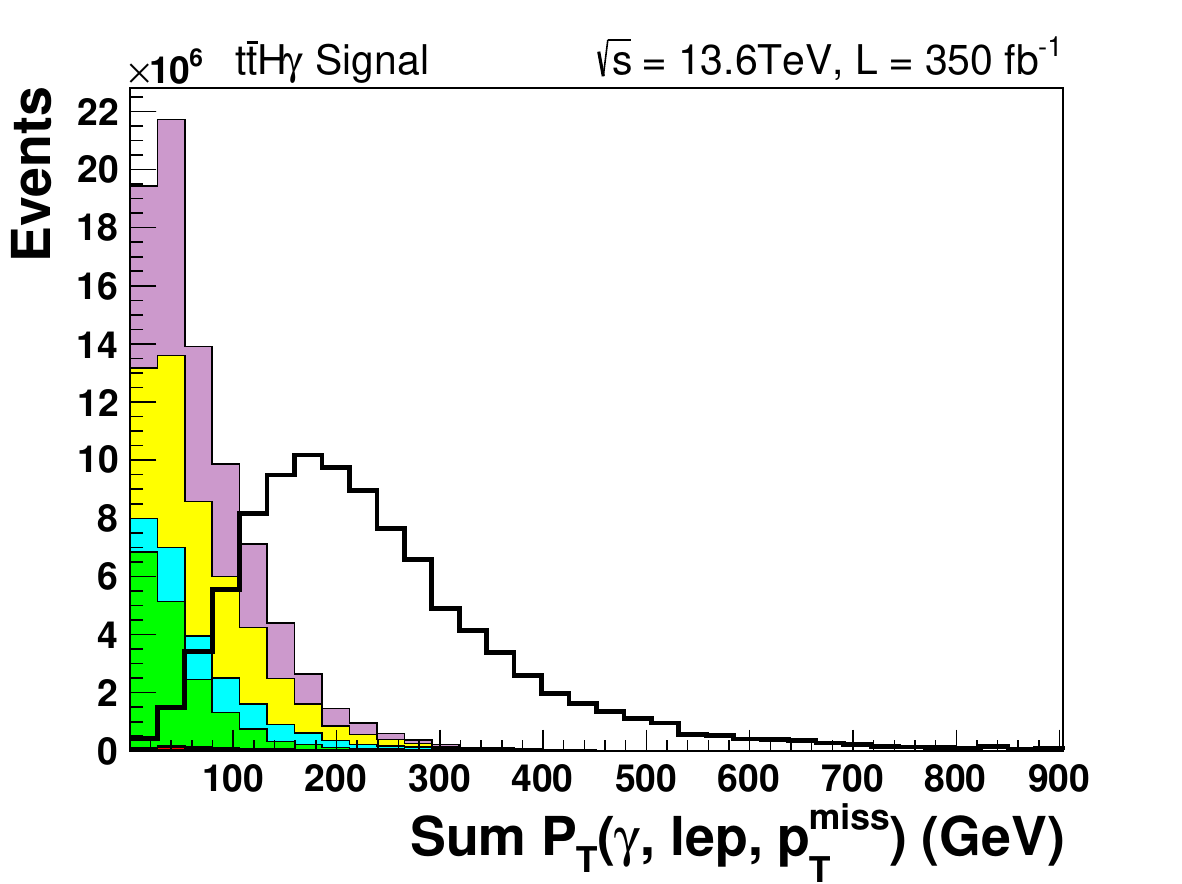}
	\end{minipage}
	
	\caption{Comparison of photon and lepton multiplicities and Sum of $p_T$ of photons, leptons and \MET\ for signal and background events. These are the best performing variables for \ttHa\ channel.}
	\label{fig:IV_ttha}
\end{figure}

\vspace{-0.26cm}
\label{sec:signal_extraction}
To differentiate signal events from background processes, Boosted Decision Trees\,(BDTs\,\cite{BDT} are utilized. The BDT technique employed is a "gradient boosting", which is accessible as part of the XGBoost library \cite{xgboost}. The ensemble is composed of a few hundred trees, each with a maximum depth of 14. The TMVA implementation of BDT is also utilized in this analysis\,\cite{TMVA}. TMVA BDT employs a similar tree-based architecture consisting of a few hundred trees, the specific BDT method used is "AdaBoost" and the maximum depth is set to 3.0. The BDTs are trained using a set of input variables that provide the most separation power between signal and background events. These variables include kinematic properties of the reconstructed objects, and b-tagging information, as previously detailed in Section \ref{sec:Simulation_Samples}. The kinematic variables used as inputs to the BDT are listed in the Table \ref{tab:bdt_appendix} and visualized in Figures \ref{fig:IV_tha} and \ref{fig:IV_ttha}.

An important aspect of the BDT is understanding which input variables contribute significantly to the classification decision. Variable importance scores are computed using XGBoost as well as TMVA, which provides insights into the relative importance of each feature in the model. The variable importance for XGBoost BDTG as well as the TMVA BDT is illustrated in Table \ref{tab:bdt_appendix}, which lists the top features used in the BDTs for both \tHa\, and \ttHa\, datasets, ranked by their importance scores. Notably, for the TMVA BDT, $\Delta\eta_{b_{12}}$ was found to contribute most significantly in the \tHa\ dataset, while the number of photons was the most important feature in the \ttHa\ dataset. For the XGBoost BDT, the most impactful variable in the \tHa\ dataset was $\Sigma p_T$ of photons, leptons, and \MET, whereas in the \ttHa\ dataset, the dominant variable was $\Sigma p_T$ of photons, leptons, jets, and \MET.

The events from signal and background are split into two subsets: one subset is employed for the training process, while the other serves as a test sample for cross validation and to monitor against overtraining. To optimize the performance of the XGBoost BDTG, we utilize a grid search approach provided by scikit-learn module \cite{sklearn}. This method helps fine-tune hyperparameters such as the learning rate, maximum depth of trees, and the number of trees, which ensures an optimal balance between model complexity and generalization.

\begin{figure}[hbtp]
	\centering
	\includegraphics[width=0.47\linewidth]{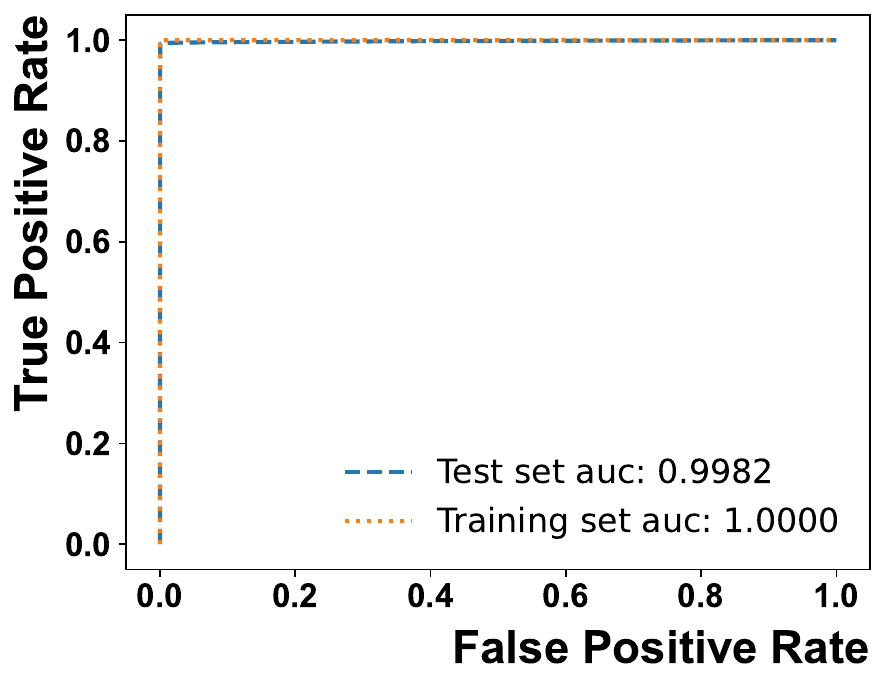}
	\hspace{0.01\linewidth}
	\includegraphics[width=0.49\linewidth]{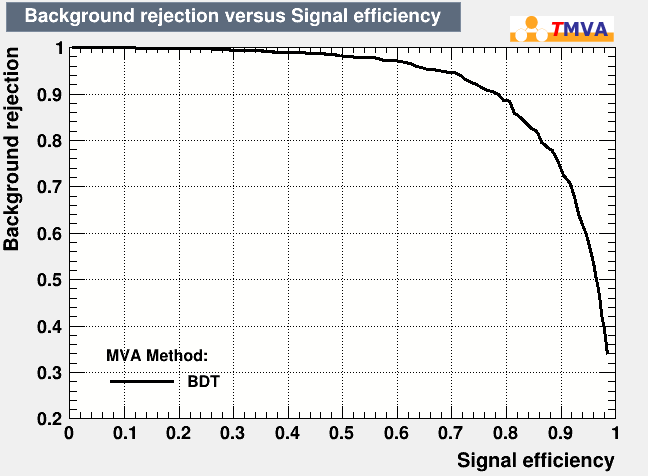}
	\caption{ROC curve for the classifier trained using \tHa\, dataset. The Area Under the Curve is shown for XGBoost BDTG (left) and TMVA BDT (right).}
	\label{fig:AUC_BDT_tha}
\end{figure}

Unlike XGBoost, which provides built-in functions from scikit-learn for hyperparameter optimization such as grid search, the tuning of TMVA BDT hyperparameters like the learning rate, maximum tree depth, and the number of trees is done manually. This manual tuning process involves iteratively running multiple training sessions, systematically varying hyperparameters, and evaluating their impact using key performance metrics such as signal significance, background rejection, and area under the Receiver Operating Characteristic (ROC) curve. The ROC curve is a graphical plot that represents the relationship between the true positive rate (signal efficiency) and the false positive rate (background efficiency) for different decision thresholds of the classifier. The area under this curve (AUC) provides a single scalar value summarizing the classifier’s ability to distinguish between signal and background, a value of 1 indicates perfect separation, while a value of 0.5 corresponds to a random guess.

\begin{figure}[hbtp]
	\centering
	\includegraphics[width=0.47\linewidth]{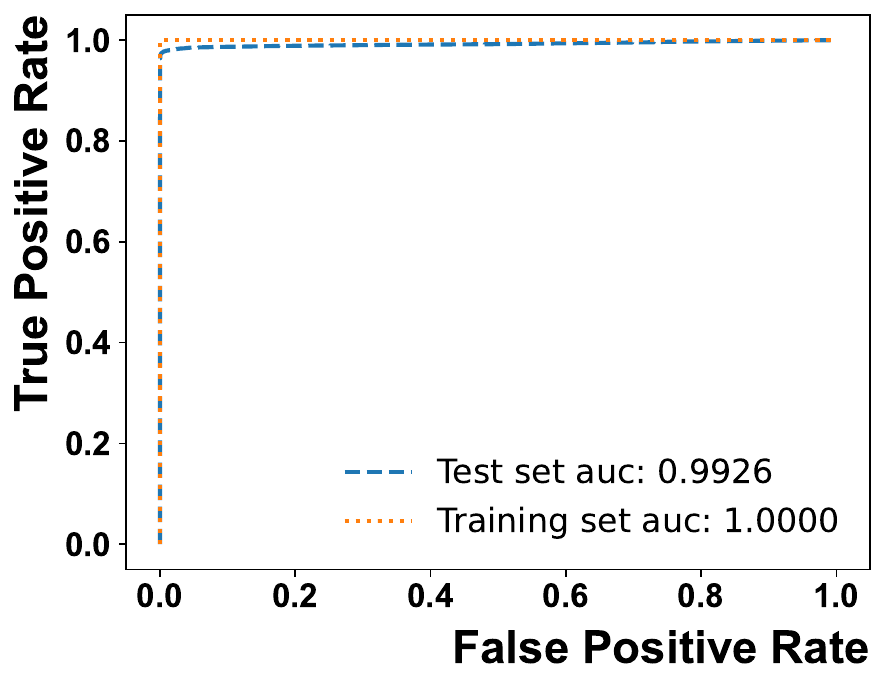}
	\hspace{0.01\linewidth}
	\includegraphics[width=0.49\linewidth]{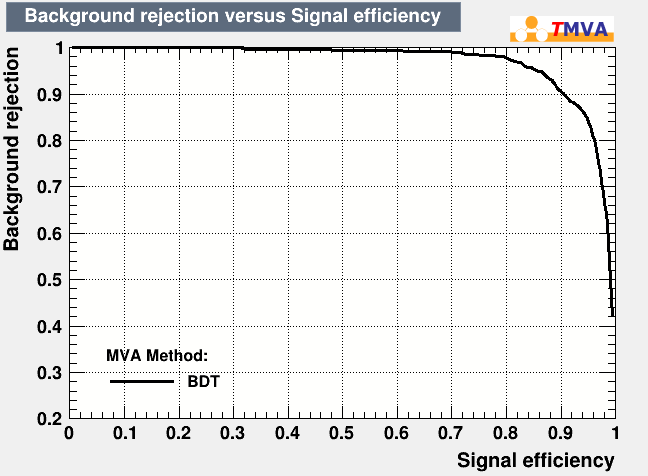}
	\caption{ROC curves for the classifier trained using the \ttHa\, dataset. Area Under the Curve is shown for XGBoost BDTG (left) and TMVA BDT (right).}
	\label{fig:AUC_BDT_ttha}
\end{figure}

In hyperparameter tuning, the number of trees is particularly important, as it directly influences the model's ability to generalize to unseen data. Once optimized, the BDT models are trained separately for \tHa\, and \ttHa\, datasets to assess their effectiveness in distinguishing signal from background.

\begin{figure}[hbtp]
	\centering
	\includegraphics[width=0.875\linewidth]{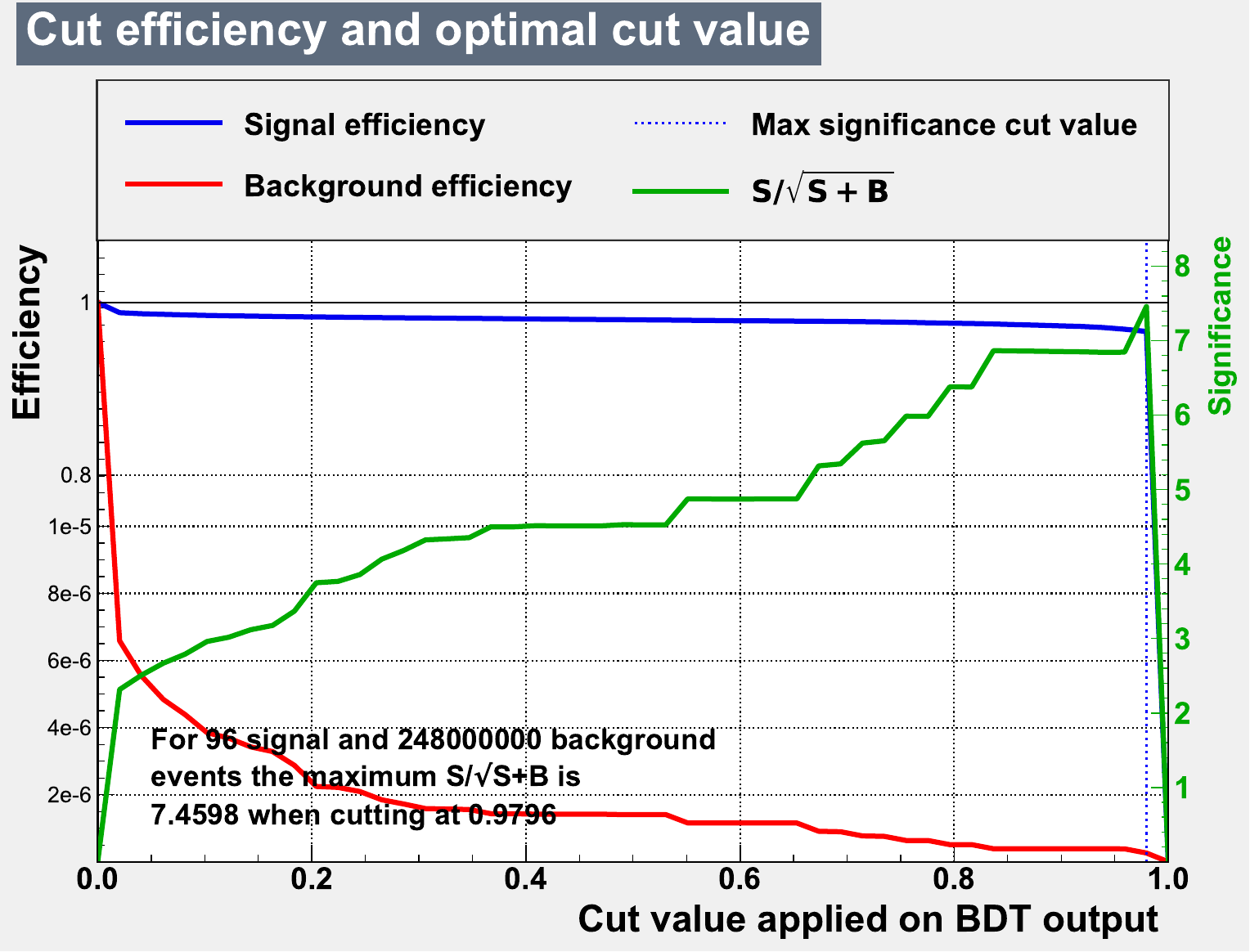}
	\caption{Efficiency and significance varying with the cut value for the \tHa\, process using XGBoost BDTG.}
	\label{fig:CCE_XGBoost_tha}
\end{figure}

To analyse the potential for overtraining of XGBoost BDT, as illustrated in Figure \ref{fig:OT_XGBoost_tha_ttha} shown in the Appendix, we monitor area under the ROC curve, as we increase the number of trees in the BDTG. Initially, as the number of trees increases, the AUC improves, indicating better classification performance. However, we observe that the AUC plateaus after a certain number of trees, suggesting that adding more trees do not significantly boost the model's performance and may lead to overfitting. The best significances (7.57 and 6.58) are achieved for somewhat complicated models with (954 and 989) trees for \tHa\, and \ttHa\, datasets respectively, however there is risk of overtraining. To prevent overtraining, we select the optimal number of trees (165 and 290) depending on the point at which the AUC stabilizes. This strategy ensures that we maintain a balance between model complexity and generalization capability.

\begin{figure}[hbtp]
	\centering
	\includegraphics[width=0.875\linewidth]{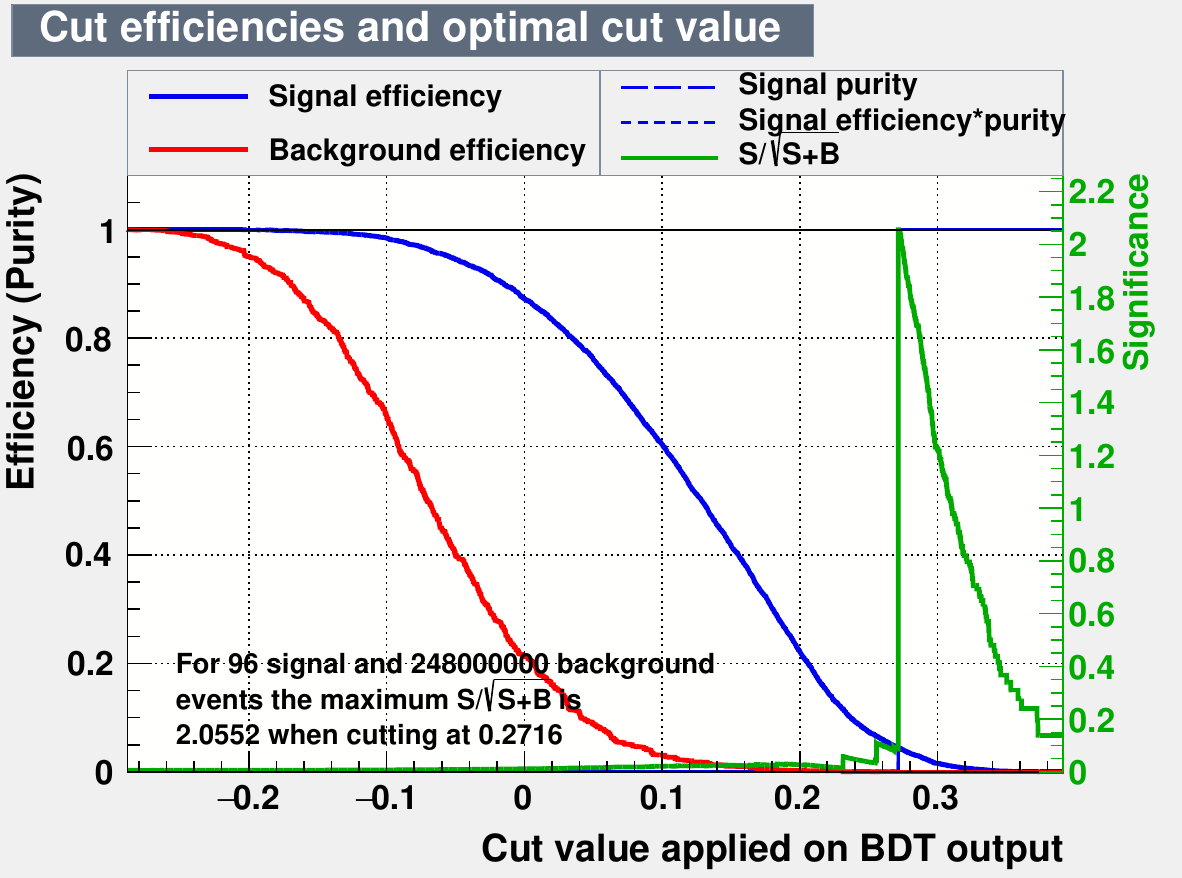}
	\caption{Efficiency and significance varying with the cut value for the \tHa\, process using TMVA BDT.}
	\label{fig:CCE_TMVA_tha}
\end{figure}

For the assessment of TMVA BDT over-training, we take a look at Figure \ref{fig:OT_TMVA_tha_ttha}. The blue and red histograms in the plot correspond to the distributions of signal and background of the test sample, respectively, whereas the overlaid markers display the training sample’s distributions. The Kolmogorov-Smirnov\,(KS) test probability value given in the plot provide a quantitative evaluation of the similarity between the training and test distributions. As discussed in \cite{KS}, a KS test p-value less than 0.05 indicates that the two samples differ significantly, implying possible overtraining. The KS test yields a p-values of 0.063 for the signal distribution and 0.827 for the background distribution for the \tHa\ dataset, while For the \ttHa\ dataset the p-values are 0.559 (signal) and 0.823 (background). The p-values in all cases are above the generally accepted threshold of 0.05, suggesting no statistically significant variation between the test and training distributions, further supporting the conclusion that over-training is improbable.

The final BDT outputs for XGBoost BDT and TMVA BDT, give improved separation between signal and background compared to any single input variable. The Area Under the Curve (AUC) plot shown in Figures \ref{fig:AUC_BDT_tha} and \ref{fig:AUC_BDT_ttha} illustrate the performance of the two classifiers in both signal channels. These results are summarized numerically in Table \ref{tab:TestAUC_Comparison}, where the AUC values for both classifiers and both signal processes are compared. While TMVA BDT demonstrates good classification performance, XGBoost BDT achieves near-optimal results, attaining the highest AUC of 0.998 for the \tHa\, process and showing superior discrimination capability between signal and background.

\begin{figure}[hbtp]
	\centering
	\includegraphics[width=0.875\linewidth]{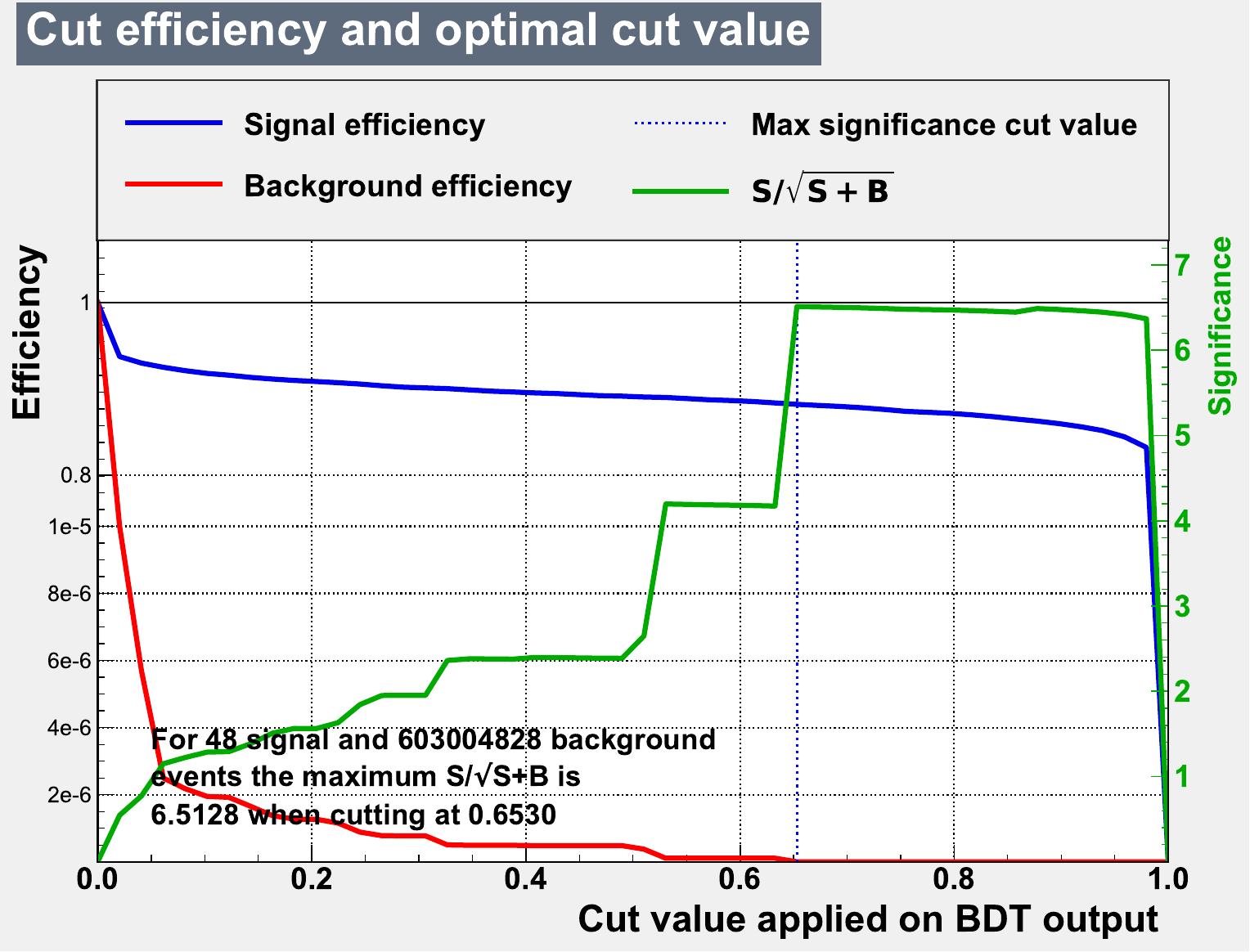}
	\caption{Efficiency and significance varying with the cut value for the \ttHa\, process using XGBoost BDTG.}
	\label{fig:CCE_XGBoost_ttha}
\end{figure}

The classification efficiency plot demonstrates the efficiency of signal and background events at various cut values applied to the BDT output. For the \tHa\ dataset, an expected event yield of 4 signal and 0 background events is obtained at the optimal cut value of 0.27 for the TMVA BDT. The optimal cut value is defined as the threshold on the BDT discriminant that maximizes the significance by achieving a balance between high signal efficiency and strong background rejection. Using these yields, a maximum significance of 2.05 is achieved, as shown in Figure \ref{fig:CCE_TMVA_tha}. In contrast, for the XGBoost BDT, an event yield of 94 signal and 65 background events is expected at the optimal cut value of 0.98, resulting in a significantly higher maximum significance of 7.46, shown in Figure \ref{fig:CCE_XGBoost_tha}.

\begin{figure}[htbp]
	\centering
	\includegraphics[width=0.875\linewidth]{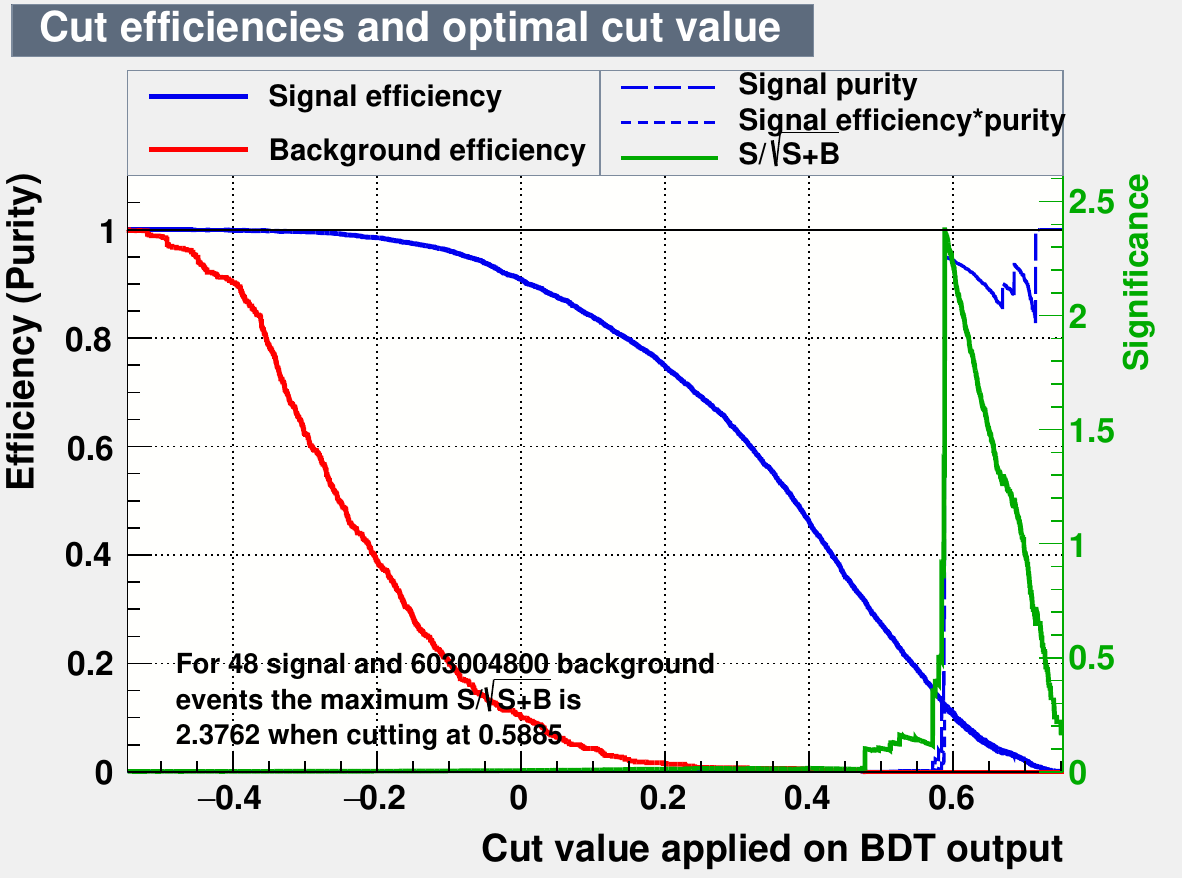}
	\caption{Efficiency and significance varying with the cut value for the \ttHa\, process using TMVA BDT.}
	\label{fig:CCE_TMVA_ttha}
\end{figure}
	
For the \ttHa\ dataset, the TMVA BDT yields 6 signal events with 0 background events at the optimal cut value of 0.59, which results in a maximum significance of 2.38, as illustrated in Figure \ref{fig:CCE_TMVA_ttha}, while 1 background and 43 signal events are expected at a cut value of 0.65 for the XGBoost BDT. Using these yields, a corresponding maximum significance of 6.51 is achieved, as presented in Figure \ref{fig:CCE_XGBoost_ttha}. The expected event yields and the corresponding significances for each model and dataset are summarized in Table \ref{tab:BDT_Results}. These values of signal and background events are rounded to the nearest integer.

\begin{table}[bht]
	\centering
	\caption{Expected signal (S) and background (B) event yields at the optimal cut value for TMVA and XGBoost classifiers on the \tHa\ and \ttHa\ datasets. Significance is calculated as S/$\sqrt{\text{S+B}}$.}
	\begin{tabular}{|c|c|c|c|c|}
		\hline
		\rule{0pt}{2.5ex}
		\multirow{2}{*}{Metric} & \multicolumn{2}{c}{\tHa} & \multicolumn{2}{c|}{\ttHa} \\
		\cline{2-5}
		\rule{0pt}{2.5ex}
		& TMVA & XGBoost & TMVA & XGBoost \\
		\hline
		\rule{0pt}{2.5ex}
		Cut Value     & 0.27 & 0.98 & 0.59 & 0.65 \\
		Signal Events     & 4.22 & 94.1 & 5.94 & 43.1 \\
		Background Events & 0 & 65 & 0.31 & 0.7 \\
		Significance    & 2.05  & 7.46  & 2.38  & 6.51 \\
		\hline
	\end{tabular}
	\label{tab:BDT_Results}
\end{table}

In these classification efficiency plots, the x-axis represents the cut value, ranging from 0 to 1. The left y-axis, represents efficiency, while the right y-axis shows the significance, calculated as  S/$\sqrt{\text{S + B}}$, where the number of events for signal and background are represented as S and B, respectively. For XGBoost, the left y-axis is divided into three segments to highlight the behaviour of signal and background efficiencies. The upper segment (0.8–1) highlights the reduction in signal efficiency with increasing cut value, the middle segment (0.00001–0.8) captures the gradual decline of both signal and background efficiencies, while the lower segment (0–0.00001) emphasizes the rapid suppression of background efficiency as it approaches zero.
\section{Summary and Conclusion}
\label{sec:Summary}
A feasibility study of the Standard Model Higgs boson production in association with a single top-quark or a top-quark pair and a photon (\tHa\, and \ttHa) is conducted, using simulated pp collision data corresponding to an integrated luminosity of 350 fb$^{-1}$ at $\sqrt{s}=13.6$ TeV. The analysis focuses on final states where each top quark decays leptonically into a charged lepton, a neutrino, and a b-quark, while the Higgs boson decays hadronically into two b-quarks. Signal events are distinguished from background processes using a Boosted Decision Tree classifier from XGBoost and TMVA, trained on kinematic and event-level variables.

The expected \tHa\, cross section is measured to be 1.31 fb, with a significance of 7.5 standard deviations from the background-only hypothesis while the expected \ttHa\, cross section is measured to be \(2.94^{+0.196}_{-0.276}~\text{fb}\), with a significance of 6.5 standard deviations from the background-only hypothesis. These results indicate that both the \tHa\ and \ttHa\ processes are potentially observable in CMS and ATLAS data, highlighting the feasibility of detecting these rare events at the LHC.

\section{Acknowledgements}
This research has been supported by the National Centre for Physics, Islamabad.

\bibliography{paper_biblio}

\begin{thebibliography}{33}%
\makeatletter
\providecommand \@ifxundefined [1]{%
 \@ifx{#1\undefined}
}%
\providecommand \@ifnum [1]{%
 \ifnum #1\expandafter \@firstoftwo
 \else \expandafter \@secondoftwo
 \fi
}%
\providecommand \@ifx [1]{%
 \ifx #1\expandafter \@firstoftwo
 \else \expandafter \@secondoftwo
 \fi
}%
\providecommand \natexlab [1]{#1}%
\providecommand \enquote  [1]{``#1''}%
\providecommand \bibnamefont  [1]{#1}%
\providecommand \bibfnamefont [1]{#1}%
\providecommand \citenamefont [1]{#1}%
\providecommand \href@noop [0]{\@secondoftwo}%
\providecommand \href [0]{\begingroup \@sanitize@url \@href}%
\providecommand \@href[1]{\@@startlink{#1}\@@href}%
\providecommand \@@href[1]{\endgroup#1\@@endlink}%
\providecommand \@sanitize@url [0]{\catcode `\\12\catcode `\$12\catcode
  `\&12\catcode `\#12\catcode `\^12\catcode `\_12\catcode `\%12\relax}%
\providecommand \@@startlink[1]{}%
\providecommand \@@endlink[0]{}%
\providecommand \url  [0]{\begingroup\@sanitize@url \@url }%
\providecommand \@url [1]{\endgroup\@href {#1}{\urlprefix }}%
\providecommand \urlprefix  [0]{URL }%
\providecommand \Eprint [0]{\href }%
\providecommand \doibase [0]{https://doi.org/}%
\providecommand \selectlanguage [0]{\@gobble}%
\providecommand \bibinfo  [0]{\@secondoftwo}%
\providecommand \bibfield  [0]{\@secondoftwo}%
\providecommand \translation [1]{[#1]}%
\providecommand \BibitemOpen [0]{}%
\providecommand \bibitemStop [0]{}%
\providecommand \bibitemNoStop [0]{.\EOS\space}%
\providecommand \EOS [0]{\spacefactor3000\relax}%
\providecommand \BibitemShut  [1]{\csname bibitem#1\endcsname}%
\let\auto@bib@innerbib\@empty
\bibitem [{\citenamefont {Chatrchyan}\ \emph {et~al.}(2012)\citenamefont
  {Chatrchyan} \emph {et~al.}}]{Higgs_CMS}%
  \BibitemOpen
  \bibfield  {author} {\bibinfo {author} {\bibfnamefont {S.}~\bibnamefont
  {Chatrchyan}} \emph {et~al.} (\bibinfo {collaboration} {CMS}),\ }\bibfield
  {title} {\bibinfo {title} {{Observation of a New Boson at a Mass of 125 GeV
  with the CMS Experiment at the LHC}},\ }\href
  {https://doi.org/10.1016/j.physletb.2012.08.021} {\bibfield  {journal}
  {\bibinfo  {journal} {Phys. Lett. B}\ }\textbf {\bibinfo {volume} {716}},\
  \bibinfo {pages} {30} (\bibinfo {year} {2012})},\ \Eprint
  {https://arxiv.org/abs/1207.7235} {arXiv:1207.7235 [hep-ex]} \BibitemShut
  {NoStop}%
\bibitem [{\citenamefont {Aad}\ \emph {et~al.}(2012)\citenamefont {Aad} \emph
  {et~al.}}]{Higgs_ATLAS}%
  \BibitemOpen
  \bibfield  {author} {\bibinfo {author} {\bibfnamefont {G.}~\bibnamefont
  {Aad}} \emph {et~al.} (\bibinfo {collaboration} {ATLAS}),\ }\bibfield
  {title} {\bibinfo {title} {{Observation of a new particle in the search for
  the Standard Model Higgs boson with the ATLAS detector at the LHC}},\ }\href
  {https://doi.org/10.1016/j.physletb.2012.08.020} {\bibfield  {journal}
  {\bibinfo  {journal} {Phys. Lett. B}\ }\textbf {\bibinfo {volume} {716}},\
  \bibinfo {pages} {1} (\bibinfo {year} {2012})},\ \Eprint
  {https://arxiv.org/abs/1207.7214} {arXiv:1207.7214 [hep-ex]} \BibitemShut
  {NoStop}%
\bibitem [{\citenamefont {Englert}\ and\ \citenamefont
  {Brout}(1964)}]{Gauge_Vector_Mesons}%
  \BibitemOpen
  \bibfield  {author} {\bibinfo {author} {\bibfnamefont {F.}~\bibnamefont
  {Englert}}\ and\ \bibinfo {author} {\bibfnamefont {R.}~\bibnamefont
  {Brout}},\ }\bibfield  {title} {\bibinfo {title} {{Broken Symmetry and the
  Mass of Gauge Vector Mesons}},\ }\href
  {https://doi.org/10.1103/PhysRevLett.13.321} {\bibfield  {journal} {\bibinfo
  {journal} {Phys. Rev. Lett.}\ }\textbf {\bibinfo {volume} {13}},\ \bibinfo
  {pages} {321} (\bibinfo {year} {1964})}\BibitemShut {NoStop}%
\bibitem [{\citenamefont {Higgs}(1964{\natexlab{a}})}]{Gauge_Fields}%
  \BibitemOpen
  \bibfield  {author} {\bibinfo {author} {\bibfnamefont {P.~W.}\ \bibnamefont
  {Higgs}},\ }\bibfield  {title} {\bibinfo {title} {{Broken symmetries,
  massless particles and gauge fields}},\ }\href
  {https://doi.org/10.1016/0031-9163(64)91136-9} {\bibfield  {journal}
  {\bibinfo  {journal} {Phys. Lett.}\ }\textbf {\bibinfo {volume} {12}},\
  \bibinfo {pages} {132} (\bibinfo {year} {1964}{\natexlab{a}})}\BibitemShut
  {NoStop}%
\bibitem [{\citenamefont {Higgs}(1964{\natexlab{b}})}]{Gauge_Bosons}%
  \BibitemOpen
  \bibfield  {author} {\bibinfo {author} {\bibfnamefont {P.~W.}\ \bibnamefont
  {Higgs}},\ }\bibfield  {title} {\bibinfo {title} {{Broken Symmetries and the
  Masses of Gauge Bosons}},\ }\href
  {https://doi.org/10.1103/PhysRevLett.13.508} {\bibfield  {journal} {\bibinfo
  {journal} {Phys. Rev. Lett.}\ }\textbf {\bibinfo {volume} {13}},\ \bibinfo
  {pages} {508} (\bibinfo {year} {1964}{\natexlab{b}})}\BibitemShut {NoStop}%
\bibitem [{\citenamefont {Guralnik}\ \emph {et~al.}(1964)\citenamefont
  {Guralnik}, \citenamefont {Hagen},\ and\ \citenamefont
  {Kibble}}]{Massless_Particles}%
  \BibitemOpen
  \bibfield  {author} {\bibinfo {author} {\bibfnamefont {G.~S.}\ \bibnamefont
  {Guralnik}}, \bibinfo {author} {\bibfnamefont {C.~R.}\ \bibnamefont
  {Hagen}},\ and\ \bibinfo {author} {\bibfnamefont {T.~W.~B.}\ \bibnamefont
  {Kibble}},\ }\bibfield  {title} {\bibinfo {title} {{Global Conservation Laws
  and Massless Particles}},\ }\href
  {https://doi.org/10.1103/PhysRevLett.13.585} {\bibfield  {journal} {\bibinfo
  {journal} {Phys. Rev. Lett.}\ }\textbf {\bibinfo {volume} {13}},\ \bibinfo
  {pages} {585} (\bibinfo {year} {1964})}\BibitemShut {NoStop}%
\bibitem [{\citenamefont {Higgs}(1966)}]{SSB}%
  \BibitemOpen
  \bibfield  {author} {\bibinfo {author} {\bibfnamefont {P.~W.}\ \bibnamefont
  {Higgs}},\ }\bibfield  {title} {\bibinfo {title} {{Spontaneous Symmetry
  Breakdown without Massless Bosons}},\ }\href
  {https://doi.org/10.1103/PhysRev.145.1156} {\bibfield  {journal} {\bibinfo
  {journal} {Phys. Rev.}\ }\textbf {\bibinfo {volume} {145}},\ \bibinfo {pages}
  {1156} (\bibinfo {year} {1966})}\BibitemShut {NoStop}%
\bibitem [{\citenamefont {Kibble}(1967)}]{SB_non_abelian}%
  \BibitemOpen
  \bibfield  {author} {\bibinfo {author} {\bibfnamefont {T.~W.~B.}\
  \bibnamefont {Kibble}},\ }\bibfield  {title} {\bibinfo {title} {{Symmetry
  breaking in nonAbelian gauge theories}},\ }\href
  {https://doi.org/10.1103/PhysRev.155.1554} {\bibfield  {journal} {\bibinfo
  {journal} {Phys. Rev.}\ }\textbf {\bibinfo {volume} {155}},\ \bibinfo {pages}
  {1554} (\bibinfo {year} {1967})}\BibitemShut {NoStop}%
\bibitem [{\citenamefont {Chatrchyan}\ \emph
  {et~al.}(2013{\natexlab{a}})\citenamefont {Chatrchyan} \emph
  {et~al.}}]{Higgs_uncer_1}%
  \BibitemOpen
  \bibfield  {author} {\bibinfo {author} {\bibfnamefont {S.}~\bibnamefont
  {Chatrchyan}} \emph {et~al.} (\bibinfo {collaboration} {CMS}),\ }\bibfield
  {title} {\bibinfo {title} {{Observation of a New Boson with Mass Near 125 GeV
  in $pp$ Collisions at $\sqrt{s}$ = 7 and 8 TeV}},\ }\href
  {https://doi.org/10.1007/JHEP06(2013)081} {\bibfield  {journal} {\bibinfo
  {journal} {JHEP}\ }\textbf {\bibinfo {volume} {06}},\ \bibinfo {pages}
  {081}},\ \Eprint {https://arxiv.org/abs/1303.4571} {arXiv:1303.4571 [hep-ex]}
  \BibitemShut {NoStop}%
\bibitem [{\citenamefont {Chatrchyan}\ \emph
  {et~al.}(2013{\natexlab{b}})\citenamefont {Chatrchyan} \emph
  {et~al.}}]{Higgs_uncer_2}%
  \BibitemOpen
  \bibfield  {author} {\bibinfo {author} {\bibfnamefont {S.}~\bibnamefont
  {Chatrchyan}} \emph {et~al.} (\bibinfo {collaboration} {CMS}),\ }\bibfield
  {title} {\bibinfo {title} {{Study of the Mass and Spin-Parity of the Higgs
  Boson Candidate Via Its Decays to Z Boson Pairs}},\ }\href
  {https://doi.org/10.1103/PhysRevLett.110.081803} {\bibfield  {journal}
  {\bibinfo  {journal} {Phys. Rev. Lett.}\ }\textbf {\bibinfo {volume} {110}},\
  \bibinfo {pages} {081803} (\bibinfo {year} {2013}{\natexlab{b}})},\ \Eprint
  {https://arxiv.org/abs/1212.6639} {arXiv:1212.6639 [hep-ex]} \BibitemShut
  {NoStop}%
\bibitem [{\citenamefont {Aad}\ \emph {et~al.}(2013)\citenamefont {Aad} \emph
  {et~al.}}]{Higgs_uncer_3}%
  \BibitemOpen
  \bibfield  {author} {\bibinfo {author} {\bibfnamefont {G.}~\bibnamefont
  {Aad}} \emph {et~al.} (\bibinfo {collaboration} {ATLAS}),\ }\bibfield
  {title} {\bibinfo {title} {{Evidence for the spin-0 nature of the Higgs boson
  using ATLAS data}},\ }\href {https://doi.org/10.1016/j.physletb.2013.08.026}
  {\bibfield  {journal} {\bibinfo  {journal} {Phys. Lett. B}\ }\textbf
  {\bibinfo {volume} {726}},\ \bibinfo {pages} {120} (\bibinfo {year}
  {2013})},\ \Eprint {https://arxiv.org/abs/1307.1432} {arXiv:1307.1432
  [hep-ex]} \BibitemShut {NoStop}%
\bibitem [{\citenamefont {Aaltonen}\ \emph {et~al.}(2013)\citenamefont
  {Aaltonen} \emph {et~al.}}]{Higgs_uncer_4}%
  \BibitemOpen
  \bibfield  {author} {\bibinfo {author} {\bibfnamefont {T.}~\bibnamefont
  {Aaltonen}} \emph {et~al.} (\bibinfo {collaboration} {CDF, D0}),\ }\bibfield
  {title} {\bibinfo {title} {{Higgs Boson Studies at the Tevatron}},\ }\href
  {https://doi.org/10.1103/PhysRevD.88.052014} {\bibfield  {journal} {\bibinfo
  {journal} {Phys. Rev. D}\ }\textbf {\bibinfo {volume} {88}},\ \bibinfo
  {pages} {052014} (\bibinfo {year} {2013})},\ \Eprint
  {https://arxiv.org/abs/1303.6346} {arXiv:1303.6346 [hep-ex]} \BibitemShut
  {NoStop}%
\bibitem [{\citenamefont {Cepeda}\ \emph {et~al.}(2019)\citenamefont {Cepeda}
  \emph {et~al.}}]{Higgs}%
  \BibitemOpen
  \bibfield  {author} {\bibinfo {author} {\bibfnamefont {M.}~\bibnamefont
  {Cepeda}} \emph {et~al.},\ }\bibfield  {title} {\bibinfo {title} {{Report
  from Working Group 2}: {Higgs Physics at the HL-LHC and HE-LHC}},\ }\href
  {https://doi.org/10.23731/CYRM-2019-007.221} {\bibfield  {journal} {\bibinfo
  {journal} {CERN Yellow Rep. Monogr.}\ }\textbf {\bibinfo {volume} {7}},\
  \bibinfo {pages} {221} (\bibinfo {year} {2019})},\ \Eprint
  {https://arxiv.org/abs/1902.00134} {arXiv:1902.00134 [hep-ph]} \BibitemShut
  {NoStop}%
\bibitem [{\citenamefont {Eichhorn}\ \emph {et~al.}(2015)\citenamefont
  {Eichhorn}, \citenamefont {Gies}, \citenamefont {Jaeckel}, \citenamefont
  {Plehn}, \citenamefont {Scherer},\ and\ \citenamefont
  {Sondenheimer}}]{Higgs_NP}%
  \BibitemOpen
  \bibfield  {author} {\bibinfo {author} {\bibfnamefont {A.}~\bibnamefont
  {Eichhorn}}, \bibinfo {author} {\bibfnamefont {H.}~\bibnamefont {Gies}},
  \bibinfo {author} {\bibfnamefont {J.}~\bibnamefont {Jaeckel}}, \bibinfo
  {author} {\bibfnamefont {T.}~\bibnamefont {Plehn}}, \bibinfo {author}
  {\bibfnamefont {M.~M.}\ \bibnamefont {Scherer}},\ and\ \bibinfo {author}
  {\bibfnamefont {R.}~\bibnamefont {Sondenheimer}},\ }\bibfield  {title}
  {\bibinfo {title} {{The Higgs Mass and the Scale of New Physics}},\ }\href
  {https://doi.org/10.1007/JHEP04(2015)022} {\bibfield  {journal} {\bibinfo
  {journal} {JHEP}\ }\textbf {\bibinfo {volume} {04}},\ \bibinfo {pages}
  {022}},\ \Eprint {https://arxiv.org/abs/1501.02812} {arXiv:1501.02812
  [hep-ph]} \BibitemShut {NoStop}%
\bibitem [{\citenamefont {Alwall}\ \emph {et~al.}(2014)\citenamefont {Alwall},
  \citenamefont {Frederix}, \citenamefont {Frixione}, \citenamefont {Hirschi},
  \citenamefont {Maltoni}, \citenamefont {Mattelaer}, \citenamefont {Shao},
  \citenamefont {Stelzer}, \citenamefont {Torrielli},\ and\ \citenamefont
  {Zaro}}]{madgraph}%
  \BibitemOpen
  \bibfield  {author} {\bibinfo {author} {\bibfnamefont {J.}~\bibnamefont
  {Alwall}}, \bibinfo {author} {\bibfnamefont {R.}~\bibnamefont {Frederix}},
  \bibinfo {author} {\bibfnamefont {S.}~\bibnamefont {Frixione}}, \bibinfo
  {author} {\bibfnamefont {V.}~\bibnamefont {Hirschi}}, \bibinfo {author}
  {\bibfnamefont {F.}~\bibnamefont {Maltoni}}, \bibinfo {author} {\bibfnamefont
  {O.}~\bibnamefont {Mattelaer}}, \bibinfo {author} {\bibfnamefont {H.-S.}\
  \bibnamefont {Shao}}, \bibinfo {author} {\bibfnamefont {T.}~\bibnamefont
  {Stelzer}}, \bibinfo {author} {\bibfnamefont {P.}~\bibnamefont {Torrielli}},\
  and\ \bibinfo {author} {\bibfnamefont {M.}~\bibnamefont {Zaro}},\ }\bibfield
  {title} {\bibinfo {title} {The automated computation of tree-level and
  next-to-leading order differential cross sections, and their matching to
  parton shower simulations},\ }\bibfield  {journal} {\bibinfo  {journal}
  {Journal of High Energy Physics}\ }\textbf {\bibinfo {volume} {2014}},\ \href
  {https://doi.org/10.1007/jhep07(2014)079} {10.1007/jhep07(2014)079} (\bibinfo
  {year} {2014}),\ \bibinfo {note} {doi:10.1007/jhep07(2014)079}\BibitemShut
  {NoStop}%
\bibitem [{\citenamefont {Arkani-Hamed}\ \emph {et~al.}(2002)\citenamefont
  {Arkani-Hamed}, \citenamefont {Cohen}, \citenamefont {Katz},\ and\
  \citenamefont {Nelson}}]{heavytop2}%
  \BibitemOpen
  \bibfield  {author} {\bibinfo {author} {\bibfnamefont {N.}~\bibnamefont
  {Arkani-Hamed}}, \bibinfo {author} {\bibfnamefont {A.~G.}\ \bibnamefont
  {Cohen}}, \bibinfo {author} {\bibfnamefont {E.}~\bibnamefont {Katz}},\ and\
  \bibinfo {author} {\bibfnamefont {A.~E.}\ \bibnamefont {Nelson}},\ }\bibfield
   {title} {\bibinfo {title} {The littlest higgs},\ }\href
  {https://doi.org/10.1088/1126-6708/2002/07/034} {\bibfield  {journal}
  {\bibinfo  {journal} {Journal of High Energy Physics}\ }\textbf {\bibinfo
  {volume} {2002}},\ \bibinfo {pages} {034–034} (\bibinfo {year} {2002})},\
  \bibinfo {note} {doi:10.1088/1126-6708/2002/07/034}\BibitemShut {NoStop}%
\bibitem [{\citenamefont {Contino}\ \emph {et~al.}(2007)\citenamefont
  {Contino}, \citenamefont {Da~Rold},\ and\ \citenamefont
  {Pomarol}}]{heavytop3}%
  \BibitemOpen
  \bibfield  {author} {\bibinfo {author} {\bibfnamefont {R.}~\bibnamefont
  {Contino}}, \bibinfo {author} {\bibfnamefont {L.}~\bibnamefont {Da~Rold}},\
  and\ \bibinfo {author} {\bibfnamefont {A.}~\bibnamefont {Pomarol}},\
  }\bibfield  {title} {\bibinfo {title} {Light custodians in natural composite
  higgs models},\ }\bibfield  {journal} {\bibinfo  {journal} {Physical Review
  D}\ }\textbf {\bibinfo {volume} {75}},\ \href
  {https://doi.org/10.1103/physrevd.75.055014} {10.1103/physrevd.75.055014}
  (\bibinfo {year} {2007}),\ \bibinfo {note}
  {doi10.1103/physrevd.75.055014}\BibitemShut {NoStop}%
\bibitem [{\citenamefont {de~Favereau}\ \emph {et~al.}(2014)\citenamefont
  {de~Favereau}, \citenamefont {Delaere}, \citenamefont {Demin}, \citenamefont
  {Giammanco}, \citenamefont {Lemaître}, \citenamefont {Mertens},\ and\
  \citenamefont {Selvaggi}}]{delphes}%
  \BibitemOpen
  \bibfield  {author} {\bibinfo {author} {\bibfnamefont {J.}~\bibnamefont
  {de~Favereau}}, \bibinfo {author} {\bibfnamefont {C.}~\bibnamefont
  {Delaere}}, \bibinfo {author} {\bibfnamefont {P.}~\bibnamefont {Demin}},
  \bibinfo {author} {\bibfnamefont {A.}~\bibnamefont {Giammanco}}, \bibinfo
  {author} {\bibfnamefont {V.}~\bibnamefont {Lemaître}}, \bibinfo {author}
  {\bibfnamefont {A.}~\bibnamefont {Mertens}},\ and\ \bibinfo {author}
  {\bibfnamefont {M.}~\bibnamefont {Selvaggi}},\ }\bibfield  {title} {\bibinfo
  {title} {Delphes 3: a modular framework for fast simulation of a generic
  collider experiment},\ }\bibfield  {journal} {\bibinfo  {journal} {Journal of
  High Energy Physics}\ }\textbf {\bibinfo {volume} {2014}},\ \href
  {https://doi.org/10.1007/jhep02(2014)057} {10.1007/jhep02(2014)057} (\bibinfo
  {year} {2014}),\ \bibinfo {note} {doi:10.1007/jhep02(2014)057}\BibitemShut
  {NoStop}%
\bibitem [{\citenamefont {Bierlich}\ \emph {et~al.}(2022)\citenamefont
  {Bierlich}, \citenamefont {Chakraborty}, \citenamefont {Desai}, \citenamefont
  {Gellersen}, \citenamefont {Helenius}, \citenamefont {Ilten}, \citenamefont
  {Lönnblad}, \citenamefont {Mrenna}, \citenamefont {Prestel}, \citenamefont
  {Preuss}, \citenamefont {Sjöstrand}, \citenamefont {Skands}, \citenamefont
  {Utheim},\ and\ \citenamefont {Verheyen}}]{pythia8.3}%
  \BibitemOpen
  \bibfield  {author} {\bibinfo {author} {\bibfnamefont {C.}~\bibnamefont
  {Bierlich}}, \bibinfo {author} {\bibfnamefont {S.}~\bibnamefont
  {Chakraborty}}, \bibinfo {author} {\bibfnamefont {N.}~\bibnamefont {Desai}},
  \bibinfo {author} {\bibfnamefont {L.}~\bibnamefont {Gellersen}}, \bibinfo
  {author} {\bibfnamefont {I.}~\bibnamefont {Helenius}}, \bibinfo {author}
  {\bibfnamefont {P.}~\bibnamefont {Ilten}}, \bibinfo {author} {\bibfnamefont
  {L.}~\bibnamefont {Lönnblad}}, \bibinfo {author} {\bibfnamefont
  {S.}~\bibnamefont {Mrenna}}, \bibinfo {author} {\bibfnamefont
  {S.}~\bibnamefont {Prestel}}, \bibinfo {author} {\bibfnamefont {C.~T.}\
  \bibnamefont {Preuss}}, \bibinfo {author} {\bibfnamefont {T.}~\bibnamefont
  {Sjöstrand}}, \bibinfo {author} {\bibfnamefont {P.}~\bibnamefont {Skands}},
  \bibinfo {author} {\bibfnamefont {M.}~\bibnamefont {Utheim}},\ and\ \bibinfo
  {author} {\bibfnamefont {R.}~\bibnamefont {Verheyen}},\ }\href@noop {}
  {\bibinfo {title} {A comprehensive guide to the physics and usage of pythia
  8.3}} (\bibinfo {year} {2022}),\ \Eprint {https://arxiv.org/abs/2203.11601}
  {arXiv:2203.11601 [hep-ph]} \BibitemShut {NoStop}%
\bibitem [{\citenamefont {Buckley}\ \emph {et~al.}(2015)\citenamefont
  {Buckley}, \citenamefont {Ferrando}, \citenamefont {Lloyd}, \citenamefont
  {Nordström}, \citenamefont {Page}, \citenamefont {Rüfenacht}, \citenamefont
  {Schönherr},\ and\ \citenamefont {Watt}}]{lhapdf6}%
  \BibitemOpen
  \bibfield  {author} {\bibinfo {author} {\bibfnamefont {A.}~\bibnamefont
  {Buckley}}, \bibinfo {author} {\bibfnamefont {J.}~\bibnamefont {Ferrando}},
  \bibinfo {author} {\bibfnamefont {S.}~\bibnamefont {Lloyd}}, \bibinfo
  {author} {\bibfnamefont {K.}~\bibnamefont {Nordström}}, \bibinfo {author}
  {\bibfnamefont {B.}~\bibnamefont {Page}}, \bibinfo {author} {\bibfnamefont
  {M.}~\bibnamefont {Rüfenacht}}, \bibinfo {author} {\bibfnamefont
  {M.}~\bibnamefont {Schönherr}},\ and\ \bibinfo {author} {\bibfnamefont
  {G.}~\bibnamefont {Watt}},\ }\bibfield  {title} {\bibinfo {title} {Lhapdf6:
  parton density access in the lhc precision era},\ }\bibfield  {journal}
  {\bibinfo  {journal} {The European Physical Journal C}\ }\textbf {\bibinfo
  {volume} {75}},\ \href {https://doi.org/10.1140/epjc/s10052-015-3318-8}
  {10.1140/epjc/s10052-015-3318-8} (\bibinfo {year} {2015}),\ \bibinfo {note}
  {doi:10.1140/epjc/s10052-015-3318-8}\BibitemShut {NoStop}%
\bibitem [{\citenamefont {Sirunyan}\ \emph {et~al.}(2017)\citenamefont
  {Sirunyan} \emph {et~al.}}]{PF_reco}%
  \BibitemOpen
  \bibfield  {author} {\bibinfo {author} {\bibfnamefont {A.~M.}\ \bibnamefont
  {Sirunyan}} \emph {et~al.} (\bibinfo {collaboration} {CMS}),\ }\bibfield
  {title} {\bibinfo {title} {{Particle-flow reconstruction and global event
  description with the CMS detector}},\ }\href
  {https://doi.org/10.1088/1748-0221/12/10/P10003} {\bibfield  {journal}
  {\bibinfo  {journal} {JINST}\ }\textbf {\bibinfo {volume} {12}}\bibfield
  {number} {\bibinfo  {number} { (10)},\ \bibinfo {pages} {P10003}},\ }\bibinfo
  {note} {doi:10.1088/1748-0221/12/10/P10003},\ \Eprint
  {https://arxiv.org/abs/1706.04965} {arXiv:1706.04965 [physics.ins-det]}
  \BibitemShut {NoStop}%
\bibitem [{\citenamefont {Beaudette}(2014)}]{PF2}%
  \BibitemOpen
  \bibfield  {author} {\bibinfo {author} {\bibfnamefont {F.}~\bibnamefont
  {Beaudette}},\ }\href@noop {} {\bibinfo {title} {The cms particle flow
  algorithm}} (\bibinfo {year} {2014}),\ \Eprint
  {https://arxiv.org/abs/1401.8155} {arXiv:1401.8155 [hep-ex]} \BibitemShut
  {NoStop}%
\bibitem [{PF1(2010)}]{PF1}%
  \BibitemOpen
  \href@noop {} {\bibinfo {title} {{Commissioning of the Particle-flow Event
  Reconstruction with the first LHC collisions recorded in the CMS
  detector}}},\ \bibinfo {howpublished}
  {\url{https://cds.cern.ch/record/1247373}} (\bibinfo {year}
  {2010})\BibitemShut {NoStop}%
\bibitem [{\citenamefont {Khachatryan}\ \emph {et~al.}(2014)\citenamefont
  {Khachatryan} \emph {et~al.}}]{diphoton}%
  \BibitemOpen
  \bibfield  {author} {\bibinfo {author} {\bibfnamefont {V.}~\bibnamefont
  {Khachatryan}} \emph {et~al.} (\bibinfo {collaboration} {CMS}),\ }\bibfield
  {title} {\bibinfo {title} {{Observation of the Diphoton Decay of the Higgs
  Boson and Measurement of Its Properties}},\ }\href
  {https://doi.org/10.1140/epjc/s10052-014-3076-z} {\bibfield  {journal}
  {\bibinfo  {journal} {Eur. Phys. J. C}\ }\textbf {\bibinfo {volume} {74}},\
  \bibinfo {pages} {3076} (\bibinfo {year} {2014})},\ \bibinfo {note}
  {doi:10.1140/epjc/s10052-014-3076-z},\ \Eprint
  {https://arxiv.org/abs/1407.0558} {arXiv:1407.0558 [hep-ex]} \BibitemShut
  {NoStop}%
\bibitem [{\citenamefont {Adam}\ \emph {et~al.}(2005)\citenamefont {Adam},
  \citenamefont {Frühwirth}, \citenamefont {Strandlie},\ and\ \citenamefont
  {Todorov}}]{e_reco1}%
  \BibitemOpen
  \bibfield  {author} {\bibinfo {author} {\bibfnamefont {W.}~\bibnamefont
  {Adam}}, \bibinfo {author} {\bibfnamefont {R.}~\bibnamefont {Frühwirth}},
  \bibinfo {author} {\bibfnamefont {A.}~\bibnamefont {Strandlie}},\ and\
  \bibinfo {author} {\bibfnamefont {T.}~\bibnamefont {Todorov}},\ }\bibfield
  {title} {\bibinfo {title} {Reconstruction of electrons with the gaussian-sum
  filter in the cms tracker at the lhc},\ }\href
  {https://doi.org/10.1088/0954-3899/31/9/n01} {\bibfield  {journal} {\bibinfo
  {journal} {Journal of Physics G: Nuclear and Particle Physics}\ }\textbf
  {\bibinfo {volume} {31}},\ \bibinfo {pages} {N9–N20} (\bibinfo {year}
  {2005})},\ \bibinfo {note} {doi:10.1088/0954-3899/31/9/n01}\BibitemShut
  {NoStop}%
\bibitem [{\citenamefont {Collaboration}(2021)}]{e_gamma_reco1}%
  \BibitemOpen
  \bibfield  {author} {\bibinfo {author} {\bibfnamefont {T.~C.}\ \bibnamefont
  {Collaboration}},\ }\bibfield  {title} {\bibinfo {title} {Electron and photon
  reconstruction and identification with the cms experiment at the cern lhc},\
  }\href {https://doi.org/10.1088/1748-0221/16/05/p05014} {\bibfield  {journal}
  {\bibinfo  {journal} {Journal of Instrumentation}\ }\textbf {\bibinfo
  {volume} {16}}\bibfield  {number} {\bibinfo  {number} { (05)},\ \bibinfo
  {pages} {P05014}},\ }\bibinfo {note}
  {doi:10.1088/1748-0221/16/05/p05014}\BibitemShut {NoStop}%
\bibitem [{\citenamefont {Rembser}(2019)}]{e_gamma_reco2}%
  \BibitemOpen
  \bibfield  {author} {\bibinfo {author} {\bibfnamefont {J.}~\bibnamefont
  {Rembser}},\ }\bibfield  {title} {\bibinfo {title} {Cms electron and photon
  performance at 13 tev},\ }\href
  {https://doi.org/10.1088/1742-6596/1162/1/012008} {\bibfield  {journal}
  {\bibinfo  {journal} {Journal of Physics: Conference Series}\ }\textbf
  {\bibinfo {volume} {1162}},\ \bibinfo {pages} {012008} (\bibinfo {year}
  {2019})},\ \bibinfo {note} {doi:10.1088/1742-6596/1162/1/012008}\BibitemShut
  {NoStop}%
\bibitem [{\citenamefont {Sirunyan}\ \emph {et~al.}(2018)\citenamefont
  {Sirunyan} \emph {et~al.}}]{PF_muon}%
  \BibitemOpen
  \bibfield  {author} {\bibinfo {author} {\bibfnamefont {A.~M.}\ \bibnamefont
  {Sirunyan}} \emph {et~al.} (\bibinfo {collaboration} {CMS}),\ }\bibfield
  {title} {\bibinfo {title} {{Performance of the CMS muon detector and muon
  reconstruction with proton-proton collisions at $\sqrt{s}=$ 13 TeV}},\ }\href
  {https://doi.org/10.1088/1748-0221/13/06/P06015} {\bibfield  {journal}
  {\bibinfo  {journal} {JINST}\ }\textbf {\bibinfo {volume} {13}}\bibfield
  {number} {\bibinfo  {number} { (06)},\ \bibinfo {pages} {P06015}},\ }\bibinfo
  {note} {doi:10.1088/1748-0221/13/06/P06015},\ \Eprint
  {https://arxiv.org/abs/1804.04528} {arXiv:1804.04528 [physics.ins-det]}
  \BibitemShut {NoStop}%
\bibitem [{\citenamefont {Coadou}(2022)}]{BDT}%
  \BibitemOpen
  \bibfield  {author} {\bibinfo {author} {\bibfnamefont {Y.}~\bibnamefont
  {Coadou}},\ }\bibinfo {title} {Boosted decision trees},\ in\ \href
  {https://doi.org/10.1142/9789811234033_0002} {\emph {\bibinfo {booktitle}
  {Artificial Intelligence for High Energy Physics}}}\ (\bibinfo  {publisher}
  {WORLD SCIENTIFIC},\ \bibinfo {year} {2022})\ Chap.\ \bibinfo {chapter}
  {Chapter 2}, p.\ \bibinfo {pages} {9–58}\BibitemShut {NoStop}%
\bibitem [{\citenamefont {Chen}\ and\ \citenamefont
  {Guestrin}(2016)}]{xgboost}%
  \BibitemOpen
  \bibfield  {author} {\bibinfo {author} {\bibfnamefont {T.}~\bibnamefont
  {Chen}}\ and\ \bibinfo {author} {\bibfnamefont {C.}~\bibnamefont
  {Guestrin}},\ }\bibfield  {title} {\bibinfo {title} {Xgboost: A scalable tree
  boosting system},\ }in\ \href {https://doi.org/10.1145/2939672.2939785}
  {\emph {\bibinfo {booktitle} {Proceedings of the 22nd ACM SIGKDD
  International Conference on Knowledge Discovery and Data Mining}}},\ \bibinfo
  {series and number} {KDD '16}\ (\bibinfo  {publisher} {Association for
  Computing Machinery},\ \bibinfo {address} {New York, NY, USA},\ \bibinfo
  {year} {2016})\ p.\ \bibinfo {pages} {785–794}\BibitemShut {NoStop}%
\bibitem [{\citenamefont {Hocker}\ \emph {et~al.}(2007)\citenamefont {Hocker}
  \emph {et~al.}}]{TMVA}%
  \BibitemOpen
  \bibfield  {author} {\bibinfo {author} {\bibfnamefont {A.}~\bibnamefont
  {Hocker}} \emph {et~al.} (\bibinfo {collaboration} {TMVA}),\ }\bibfield
  {title} {\bibinfo {title} {{TMVA - Toolkit for Multivariate Data Analysis}},\
  }\href@noop {} {\  (\bibinfo {year} {2007})},\ \Eprint
  {https://arxiv.org/abs/physics/0703039} {arXiv:physics/0703039} \BibitemShut
  {NoStop}%
\bibitem [{\citenamefont {Pedregosa}\ \emph {et~al.}(2011)\citenamefont
  {Pedregosa} \emph {et~al.}}]{sklearn}%
  \BibitemOpen
  \bibfield  {author} {\bibinfo {author} {\bibfnamefont {F.}~\bibnamefont
  {Pedregosa}} \emph {et~al.},\ }\bibfield  {title} {\bibinfo {title}
  {{Scikit-learn: Machine Learning in Python}},\ }\href@noop {} {\bibfield
  {journal} {\bibinfo  {journal} {J. Machine Learning Res.}\ }\textbf {\bibinfo
  {volume} {12}},\ \bibinfo {pages} {2825} (\bibinfo {year} {2011})},\ \Eprint
  {https://arxiv.org/abs/1201.0490} {arXiv:1201.0490 [cs.LG]} \BibitemShut
  {NoStop}%
\bibitem [{\citenamefont {Demortier}\ and\ \citenamefont {Lyons}(2014)}]{KS}%
  \BibitemOpen
  \bibfield  {author} {\bibinfo {author} {\bibfnamefont {L.}~\bibnamefont
  {Demortier}}\ and\ \bibinfo {author} {\bibfnamefont {L.}~\bibnamefont
  {Lyons}},\ }\bibfield  {title} {\bibinfo {title} {Testing hypotheses in
  particle physics: Plots of $p_{0}$ versus $p_{1}$},\ }\href@noop {} {\
  (\bibinfo {year} {2014})},\ \Eprint {https://arxiv.org/abs/1408.6123}
  {arXiv:1408.6123} \BibitemShut {NoStop}%
\end{thebibliography}%

\appendix
\onecolumngrid

\section*{Appendix}
\label{appendix}
\begin{figure}[H]
	\centering
	\includegraphics[width=0.44\linewidth]{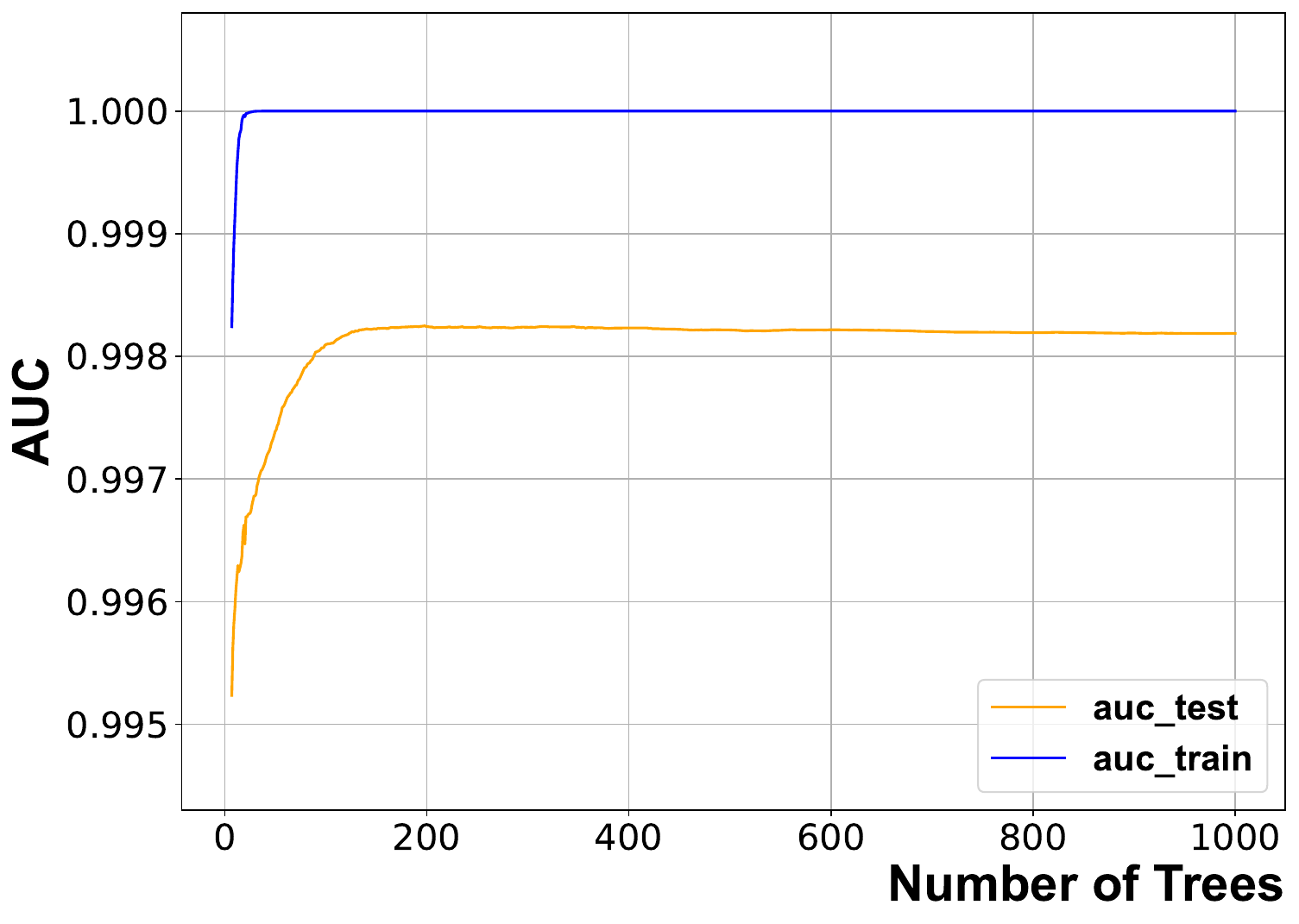}
	\hspace{0.04\linewidth}
	\includegraphics[width=0.44\linewidth]{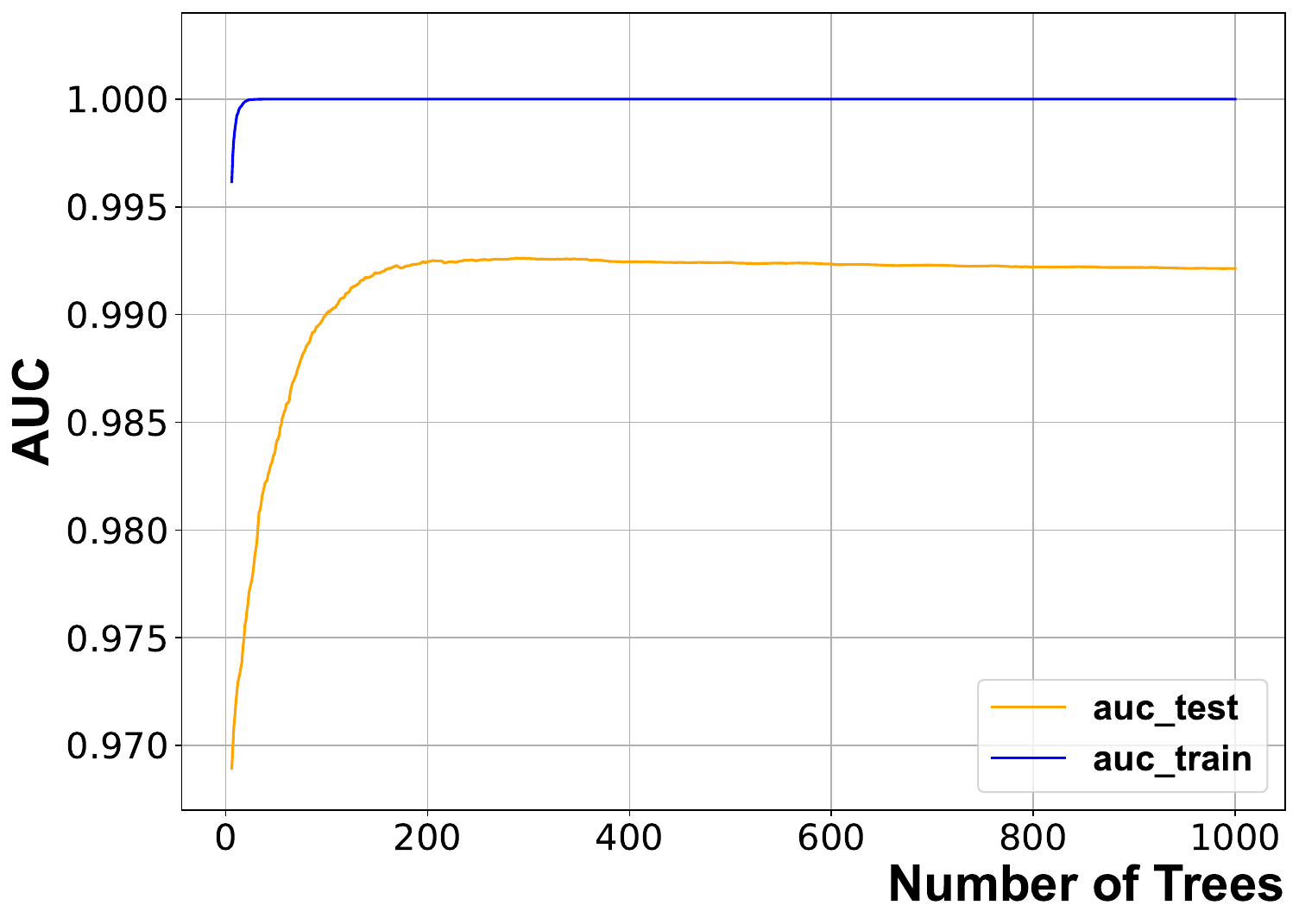}
	\caption{AUC for the XGBoost BDTG varying with number of trees for the \tHa\,process\,(left), \ttHa\,process\,(right).}
	\label{fig:OT_XGBoost_tha_ttha}
\end{figure}
\twocolumngrid

\end{document}